\documentclass[%
preprint,onecolumn,
 amsmath,amssymb,
 aps,
]{revtex4-1}
\usepackage{natbib}
\usepackage{subfigure}
\usepackage{amsthm}
\usepackage{soul} %for crossing sentences.
\newtheorem{theorem}{Theorem}
\newtheorem{lemma}{Lemma}
\newtheorem{proposition}{Proposition}

\newenvironment{definition}[1][Definition]{\begin{trivlist}
\item[\hskip \labelsep {\bfseries #1}]}{\end{trivlist}}

\newenvironment{remark}[1][Remark]{\begin{trivlist}
\item[\hskip \labelsep {\bfseries #1}]}{\end{trivlist}}
\usepackage{graphicx}% Include figure files
\usepackage{dcolumn}% Align table columns on decimal point
\usepackage{bm}% bold math
\usepackage{bbm} % for complex iota
\usepackage{mathtools} % for colonequal
\usepackage{color}
\begin{document}

%\preprint{APS/123-QED}

\title{
Dynamical aspects of a restricted three-vortex problem}
%On the dynamics of a free point vortex pair in presence %of a fixed point vortex}% Force line breaks with \\

%\author{Sreethin Sreedharan K.}
%%\author{Priyanka Shukla}%
%%\email{Second.Author@institution.edu}
%\affiliation{%
%Department of Mathematics,
%Indian Institute of Technology Madras, Chennai 600036, India
%}%

\author{Sreethin Sreedharan K\footnote{sreethin2@gmail.com} and Priyanka Shukla\footnote{priyanka@iitm.ac.in}}
%\email{}

%\author{Priyanka Shukla}
%\email[(corresponding author) ]%{priyanka@iitm.ac.in}
%\homepage{https://home.iitm.ac.in/priyanka/}
\affiliation{Department of Mathematics,\\
Indian Institute of Technology Madras, Chennai 600036, India
}

\date{\today}

\begin{abstract}
The restricted three-vortex problem is investigated with one of the point vortices fixed in the plane. The motion of the free vortex having zero circulation is explored from a rotating frame of reference within which the free vortex with non-zero circulation remains stationary.  By using the basic dynamical system theory, it is shown that the vortex motion is always bounded and any configuration of vortices must go through at least one  collinear state. 
%The present analysis also reveals that any non-fixed configuration of vortices  either has the  inter-vortex distances periodic in time or  is aperiodic with vortices  asymptotically approaching  a collinear fixed configuration. T
The present analysis reveals that the inter-vortex distances in any non-fixed configuration of vortices are either periodic in time or aperiodic in time with vortices asymptotically converging to a collinear fixed configuration. The initial conditions required for different 
%each of the three 
types of motion are explained in detail by exploiting the Hamiltonian structure of the problem.
\end{abstract}

\pacs{Valid PACS appear here}% PACS, the Physics and Astronomy
\maketitle

%\tableofcontents

\section{Introduction}
The point vortex model, first introduced by Helmholtz in his seminal paper~\cite{HH1858}, is the simplest and the most analytically amenable model for a vortex in an ideal two-dimensional incompressible flow. In this model,  the curl of the velocity field, namely the vorticity, is assumed to follow a set of discrete singularities, i.e.,  a superposition of delta distributions. Similarly to the $N$-body problem in celestial mechanics, the problem on the motion of a system of mutually interacting $N$ point  vortices is  called  an $N$-vortex problem. Investigating these problems is the first step towards understanding complex vortex interactions in fluid evolutions.  Further details on vorticity and $N$-vortex problems can be found in~\cite{S1992} and~\cite{NP2013}, respectively.

The solution of one- as well as two-vortex problem is rather trivial, whereas $N$-vortex problem for $N\geq 4$ is not integrable in general, and analytical solutions are only available in some special cases~(see, e.g.~\cite{A1983,NP2013} for more details). As a result of its integrability and non-trivial set of solutions, the three-vortex problem in the two-dimensional plane has been extensively studied in the literature~\cite{G1877,S1949,N1975,AH1979,H2007,AH2010,G2016,NS1979,TT1988,AH2009,KA2018,KS2018}. Most of these studies  are  qualitative in nature as the  analytic expressions of the solutions typically  involve  elliptic or hyperelliptic integrals that do not provide any insights on the actual vortex motion. In 2013,~\citet{RK2013} and later in 2018, Koshel et al.~\cite{KR2018} in a revisited paper have looked at a variant of the three-vortex problem in which one among the three vortices is assumed to be fixed at one location of the Euclidean plane for all time. Although they~\cite{RK2013,KR2018} have restricted their attention to a counter-rotating pair of vortices, their results indicate that the solutions of this model have flavors from both two- and three-vortex problems. In the numerical section of~\cite{KR2018}, the authors have illustrated several examples of vortex motion from a rotating frame of reference within which one of the free vortices is stationary. Vortex trajectories in this frame of reference  appear to be much simpler than those in a fixed reference frame. It is intriguing to formulate this insight mathematically and to explain the variant of three-vortex problem from a purely dynamical system point of view without limiting to a counter-rotating case. Similar reduction methods have widely been employed in the past for simplifying and studying different mechanical systems with symmetry (see, e.g.~\cite{MR2013,MW1974,MW2001}). To the best of authors' knowledge, such methods have been rarely used in point vortex models~\cite{KC1985,K1988,KC1989,PM1998,AB2016}, especially on $N$-vortex problems defined in an unbounded plane. 
%   The physically relevant properties like vortex entrapments, self-similar evolutions and fixed configurations is found to be easily trackable in this new setup.  The vortex entrapments results obtained by~\cite{RK2013,KR2018} for the counter rotating pair case comes out naturally  as a sub case.

%This paper is the end result of our attempt to develop the necessary concepts and finding the correct  mathematical tools to  analyse the above variant of three-vortex problem without limiting to a counter rotating case.  i.e.,  restricting ourselves to the case in which  one of the free vortices is a weak vortex having negligible strength. 

%Although our ultimate aim is to address the above variant of three-vortex problem in the most general set-up possible, in order to first develop the necessary concepts, we  restrict ourselves to the case in which  one of the free vortices is a weak vortex having negligible strength. Considering the  limit case first, we also 

%Although our ultimate aim is to address the above variant of three-vortex problem in the most general set-up possible, in this paper, we will  restrict ourselves to the limit case in which  one of the free vortices is a weak vortex having negligible strength. By considering this `restricted' problem first, we hope to develop the necessary concepts and finding the correct  mathematical tools to  analyse the general `un-restricted problem'.

 When one of the vortex circulations is zero in an $N$-vortex problem, it is referred to as a \emph{restricted $N$-vortex} problem. Our ultimate aim is to address the variant of the three-vortex problem proposed by~\citet{RK2013} in the most general possible set-up. However, in the present study, we shall focus on its restricted version, i.e., three-vortex problem in which one vortex is fixed and one vortex has zero circulation. This assumption brings forth the necessary simplicity and clarity required to develop the ideas naturally, and to understand its physical interpretations easily. We hope to generalize the present analysis and to tackle the original {\it unrestricted} problem elsewhere in the near future. 

There is a growing body of literature that concerns with vortex configurations which move as a rigid body, without change of size or shape (see, e.g.~\cite{AN2002} for a detailed review). These vortex patterns are observed in several physical situations, for instance, rotating superfluid Helium~\cite{YG1979}, magnetically confined non-neutral plasma~\cite{DF2000}, etc. Owing to the wide range of physical applications, it is worthwhile to analyze such {\it fixed vortex configurations}.  
%In our analysis, we have given special attention to these `fixed vortex configurations.' 
The overall effects that fixing a vortex has on the vortex trajectories is also investigated by comparing  the results obtained  with   that of the  classical three-vortex problem~\cite{A2009,AB2016}. As the center of vorticity~\cite{NP2013,RK2013} is no longer conserved once one fixes a vortex in the plane, such a comparison may provide some critical insights about vortex  motion having broken  symmetries.

The paper is organized as follows. In Sec.~II, we  shall mathematically formulate the problem at hand and review some of the basic concepts in the theory of dynamical system that will be used in the later sections. In Sec.~III, the underlying differential equations for the vortex coordinates in a rotating frame of reference having two of the three vortices stationary are presented, and the motion of the third vortex having zero circulation is then studied from a dynamical system point of view. Some of the special cases are explained in Sec.~IV. For the sake of completion, a short comparison is also provided on the solutions of the restricted three-vortex problem with and without a fixed vortex in the penultimate section, Sec.~V. In the last section, Sec.~VI, we summarize our findings and discuss possible future directions.
 
\section{Problem formulation}
\begin{figure}[!htpb]
\centering
\includegraphics[height=2.3in]{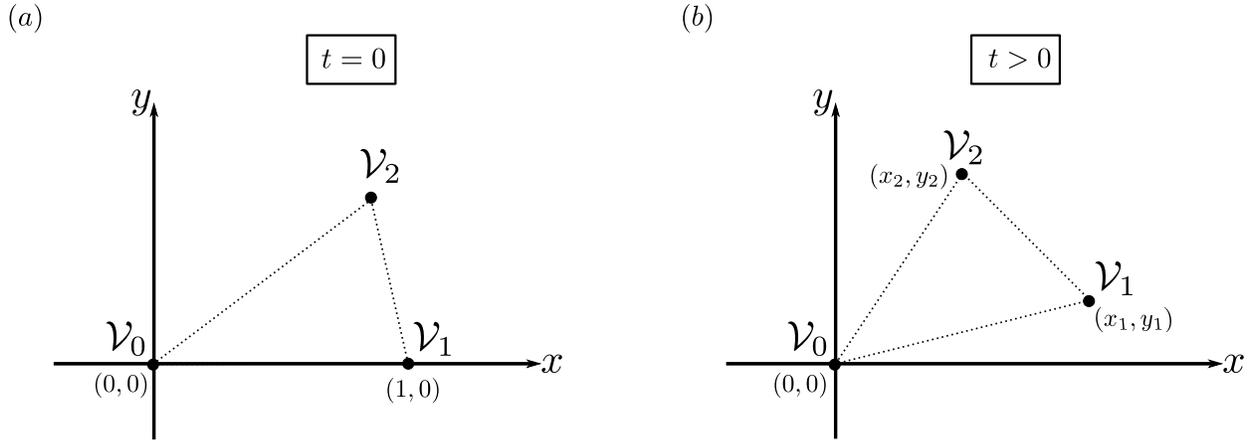}
\label{figure_model}
\caption{\small{A schematic diagram of the three-vortex model~\eqref{z0_eqn}--\eqref{z2_eqn}. (a) At time $t=0$, the vortex $\mathcal{V}_1$ is assumed to be at $(1,0)$. (b) For time $t>0$, the two free vortices $\mathcal{V}_1$ and $\mathcal{V}_2$ move freely around the fixed vortex $\mathcal{V}_0$.}}
\end{figure} 
We consider three point vortices $\{\mathcal{V}_0,\mathcal{V}_1,\mathcal{V}_2\}$ in  $\mathbb{R}^2$ plane. The restricted vortex problem, by definition, has one of the vortices having zero circulation. In the present work,  we focus ourselves to the case when one among the three vortices, say $\mathcal{V}_0$ with non-zero circulation $\Gamma_0$, is fixed at a position $(x_0,y_0)$ in $\mathbb{R}^2$, and the other two vortices, $\mathcal{V}_1$ with non-zero circulation $\Gamma_1$, and $\mathcal{V}_2$ with zero circulation, are spatially evolving  with positions $(x_1,y_1)$ and $(x_2,y_2)$ respectively. Each vortex experiences a velocity field that is a sum of the individual velocity fields produced by the  other two vortices. Without loss of generality (WLOG), one can choose a  coordinate system in such a way that the fixed vortex $\mathcal{V}_0$ is positioned at the origin, and the vortex $\mathcal{V}_1$ is initially situated along the positive $x$-axis  at a  unit distance away from the fixed vortex, by appropriately orienting the axes and choosing the length scale.
Therefore $(x_0,y_0)\equiv (0,0)$ and $(x_1,y_1)\big|_{t=0}=(1,0)$.   For the sake of simplicity, here onwards, we will identify the elements of the Cartesian plane $\mathbb{R}^2$ with that of the complex plane $\mathbb{C}$ by employing the map $(x,y)\mapsto x+\mathbbm{i}\,y$. The motion of the vortices are given by the following  autonomous  system of first-order differential equations:

\begin{align}
\dot{z_0} &=0,\label{z0_eqn}\\
\dot{z_1}&= \frac{\mathbbm{i}\Gamma_0}{2\pi}\frac{z_1}{|z_1|^2},\label{z1_eqn}\\
\dot{z_2}&= \frac{\mathbbm{i}\Gamma_0}{2\pi}\frac{z_2}{|z_2|^2}+\frac{\mathbbm{i}\Gamma_1}{2\pi}\frac{z_2-z_1}{|z_1-z_2|^2}\label{z2_eqn},
\end{align}
where $z_j(t)=x_j(t)+\mathbbm{i}\,y_j(t),\:  j\in\{0,1,2\}$ and the dot over the variables denotes the time derivative. Note that~\eqref{z1_eqn} and~\eqref{z2_eqn} are undefined   when $|z_1|=0$ and $|z_2|=0$ or $|z_2-z_1|=0$, respectively. The point vortex setting fails to explain the further evolution of vortices once such vortex collapse---vortices coalesces into a point---is encountered during the motion. 
Therefore, let us assume that initially there is no vortex collapse, which means that  $|z_1|$, $|z_2|$, and $|z_1-z_2|$ are non-zero initially. Recall that we have already assumed $|z_1|=1$ at $t=0$.

Since we are not interested in the exact motion of the vortices, but rather on the qualitative behaviour, such as the relative configurations, boundedness of vortices etc., it is  easier to look at the dynamics in the polar coordinates, i.e., $z_j(t)=r_j(t)\,e^{\mathbbm{i}\theta_j(t)}$ with $r_j(t)$ and $\theta_j(t)$ being the modulus and the argument of $z_j(t)$, respectively. From basic trigonometry,
\begin{equation}\label{cosine_rule}
r_{12}^2=r_1^2+r_2^2-2r_1r_2\cos(\theta_2-\theta_1),
\end{equation}
where $r_{12}=|z_1-z_2|$ is the distance between the vortices $\mathcal{V}_1$ and $\mathcal{V}_2$.
%From the above equations 
Let us now recall some of the terminologies associated with vortex motion and dynamical systems that will be used in the later sections.
\begin{definition}{(Fixed configuration)}
A \emph{fixed configuration} is a configuration of the three vortices  for which the vortex triangle remains fixed, i.e.,  the lengths of the three sides of the vortex triangle remain unchanged, and the motion is a rigid body  motion. A fixed configuration is said to be a \emph{fixed equilibrium} if the configuration neither rotates nor translates. Otherwise, it is called a \emph{relative equilibrium}.
\end{definition}
\begin{definition}{(Hamiltonian system)}
\label{defn_hamiltonian}
Let $H$ be a smooth real-valued function of two variables. A system of the form $\dot{x}=\partial H/\partial y$, $\dot{y}=-\partial H/\partial x$ is called a \emph{Hamiltonian system} where $H$ is called the \emph{Hamiltonian}. Note that $\dot{H}=\left(\partial H/\partial x\right)\, \dot{x}+\left(\partial H/\partial y\right)\, \dot{y}=0$. Consequently, $H(x,y)$ is a constant along any solution of the Hamiltonian system.
\end{definition}
\begin{definition}{(Reversible system)} A system of the form $\dot{x}=f(x,y), \dot{y}=g(x,y)$ is said to be \emph{reversible} if it is invariant under the transformation $t\rightarrow -t$ and $y\rightarrow -y$, i.e., if it satisfies $f(x,-y)=-f(x,y)$ and $g(x,-y)=g(x,y)$.
\end{definition}
\begin{definition}{(Index of a closed curve/equilibrium point)}    
Consider a system of the form $\dot{x}=f(x,y), \dot{y}=g(x,y)$  with $(f,g)$ being  a smooth vector field on $\mathbb{R}^2$. Let $\gamma$ be a Jordan curve (a piecewise smooth, simple closed curve) in the plane that does not have any zeros of the field (equilibrium points) on it. The \emph{index of the curve} $\gamma$ with respect to the vector field $(f,g)$ is an integer defined as the total number of anti-clockwise revolutions made by the field,  as one moves  counter-clockwise around $\gamma$ exactly once.  The above definition on the index of a closed curve can be used to define \emph{the index of an isolated  equilibrium point} $(x^*,y^*)$ as the index of any Jordan curve that contains $(x^*,y^*)$ (in its interior) and no other equilibrium points.

The index theory can be used to gain insights about the nature and number of equilibrium points of a dynamical system. The readers may refer~\cite{PL2013,SS2018} for more details about the index theory.

\end{definition}
%
%\begin{definition}(Saddle and center) Consider a non-linear system $\dot{X}=F(X)$
%\end{definition}
\begin{definition}{(Heteroclinic orbit and homoclinic orbit) }A trajectory in the phase plane which joins two distinct equilibrium points is called a \emph{heteroclinic orbit}, whereas a trajectory which joins a saddle equilibrium point to itself is called a \emph{homoclinic orbit}. 
\end{definition}
 
\section{Restricted three-vortex problem}
%Owing to the fact that the vortex $\mathcal{V}_1$ solely experiences the velocity field induced by the fixed vortex $\mathcal{V}_0$, 
From~\eqref{z1_eqn}, we find that the derivative of $|z_1|^2$ is identically zero and therefore
\begin{equation}
|z_1|^2=\text{ constant}.
\end{equation}
Since initially $|z_1|=1$, consequently $|z_1(t)|\equiv 1$ for all time $t$. Back substituting for $|z_1|$ in equation~\eqref{z1_eqn} yields $z_1(t)= e^{\mathbbm{i}wt}$, where $\omega=\Gamma_0/2\pi$. Physically, this means that 
the vortex $\mathcal{V}_1$ is simply rotating around the vortex $\mathcal{V}_0$ with a constant angular velocity $\omega$ along a  unit circle. 

To simplify the problem, we introduce a new set of co-ordinates
\begin{align}
\label{eq_coordinate_transform}
\left(\eta_0,\eta_1,\eta_2\right)\coloneqq e^{-\mathbbm{i}\omega t}\left(z_0,z_1,z_2\right)=\left(0,1,e^{-\mathbbm{i}\omega t}z_2\right).
\end{align}
The idea is to look at the motion of the vortex $\mathcal{V}_2$ from a rotating frame of reference so that only the transformed variable $\eta_2$ has any temporal evolution.
Differentiating $\eta_2$  with respect to time, we obtain
\begin{equation}
\dot{\eta_2}=-\mathbbm{i}\omega\eta_2+\frac{\mathbbm{i}\Gamma_1}{2\pi} \frac{\eta_2-\eta_1}{|\eta_2-\eta_1|^2}+\frac{\mathbbm{i}\Gamma_0}{2\pi}\frac{\eta_2}{|\eta_2|^2}.
\end{equation}
Equating the real and imaginary parts on both sides of the above expression, we find the underlying differential equations as
\begin{align}
\left.
\begin{aligned}
\dot{u}&=\omega v-\frac{\Gamma_1}{2\pi}\frac{v}{(u-1)^2+v^2}-\frac{\Gamma_0}{2\pi}\frac{v}{u^2+v^2},
%\label{udot_restricted}
\\
\dot{v}&=-\omega u+\frac{\Gamma_1}{2\pi}\frac{u-1}{(u-1)^2+v^2}+\frac{\Gamma_0}{2\pi}\frac{u}{u^2+v^2},
%\label{vdot_restricted},
\end{aligned}
\right\}
\label{udotvdot_restricted}
\end{align}
where $u$ and $v$ are the real and the imaginary parts of $\eta_2$, respectively. Since system~\eqref{udotvdot_restricted} is undefined at $(0,0)$ and $(1,0)$, we shall henceforth address them as the singularity points $\eta_0$ and $\eta_1$, which also denote the location of the vortices $\mathcal{V}_0$ and $\mathcal{V}_1$ in the complex plane, respectively. 

Note that the above system is reversible, i.e., equations~\eqref{udotvdot_restricted} are invariant under the transformation $t\rightarrow -t$ and $v\rightarrow -v$. This means that for every trajectory in the  positive $v$ plane, there is a twin trajectory in  the negative $v$ plane. They are reflections of each other along the $u$-axis but with arrows reversed. Hence, if a trajectory intersects the $u$-axis at two distinct points in finite time then it must be a closed trajectory and therefore, periodic.
%\begin{remark} 
Observe that the $(u,v)$ coordinates are related to the polar coordinates $(r,\theta)$ via
\begin{align*}
r_2^2&=u^2+v^2,\quad
r_{12}^2=(u-1)^2+v^2\quad \mbox{and}\quad
\theta_2-\theta_1=\tan^{-1}(v/u).
\end{align*}
Hence points in the $(u,v)$ phase plane, which lie on the $u$-axis ($v=0$) correspond to  collinear vortex configurations ($\theta_2=\theta_1$) in the physical plane. We shall now analyze the equilibrium points and trajectories of $\eta_2$ governed by \eqref{udotvdot_restricted}.
%\end{remark}

%--------------------------------
\subsection{Equilibrium points}
%--------------------------------

Substituting $\dot{u}=\dot{v}=0$ in  \eqref{udotvdot_restricted}, we see that all equilibrium points lie on the $u$-axis and are the real roots of the cubic polynomial 
\begin{equation}
p(u)= u^3-u^2-(\alpha+1)u+1,\quad \alpha=\Gamma_1/\Gamma_0. \label{u_fixedpoint_cubic_polynomial}
\end{equation}
%The discriminant of the above polynomial is $D(\alpha)=\alpha \,( 4 \alpha^2+13\alpha+32)$. 
Since any cubic polynomial with real coefficients has at least one real root,  at least one equilibrium point can be found on the $u$-axis. The sign of the discriminant $D(\alpha)=\alpha \,( 4 \alpha^2+13\alpha+32)$ determines the number of real roots of the polynomial $p$. As $D(0)=0$ and $D'(\alpha)=12\left(\alpha+13/12\right)^2+215/12>0$, $D(\alpha)$ and $\alpha$ have the same sign. Consequently, if the circulations $\Gamma_0$ and $\Gamma_1$ have opposite signs ($\alpha<0$), there will only be one equilibrium  point (as $D(\alpha)<0$ implies only one real root for $p$) and if they have the same sign ($\alpha>0$), there will be exactly three distinct equilibrium points on the $u$-axis (as $D(\alpha)>0$ implies $p$ has three distinct real roots). Note that $p(0)=1$ and $p(1)=-\alpha\neq 0$, this implies that $u=0$ and $u=1$ cannot  satisfy $p(u)=0$. Thus, none of the roots of the polynomial $p$ can be the singularity points $\eta_0$ and $\eta_1$ of system~\eqref{udotvdot_restricted}.

Since, the cubic term dominates for large values of $|u|$, $p(u)$ is negative for large negative values of $u$ and  positive for  large positive values of $u$. Hence, from the intermediate value theorem, it is concluded that for the case of $\alpha<0$, the only real root must lie in the interval $ (-\infty,0)$ and for the case of $\alpha>0$, the three real roots must lie in the intervals $(-\infty,0),(0,1)$ and $(1,\infty)$, respectively.

As all the equilibrium points have their $v$-component zero, the Jacobian matrix associated with~\eqref{udotvdot_restricted}, when evaluated at the equilibrium points have both the diagonal entries zero, and the characteristic polynomial of the matrix turns out to be of the  form $\lambda^2+c=0$, with $c\neq 0$.  Consequently, for the linearized system corresponding to system~\eqref{udotvdot_restricted}, an equilibrium point is either a center or a saddle point. Since a saddle is a hyperbolic equilibrium point, a saddle point of the linearized system remains a saddle with respect to the original non-linear system~\cite{PL2013}. The reversibility of system~\eqref{udotvdot_restricted} also guarantees that a linear center remains a center. Therefore, we conclude that an equilibrium point of the non-linear system~\eqref{udotvdot_restricted} is either a saddle point or a center.

%From Hartman-Grobman theorem (see, e.g.~\cite{PL2013}), it follows that any equilibrium point which is a saddle point with respect to the linearized system will remain a saddle with respect to the original non-linear system. Similarly, the reversibility of system~\eqref{udotvdot_restricted} guarantees  that a linear center is also a n 

%Since a saddle point is a hyperbolic equilibrium point, a saddle point of the linearized system will remain a saddle for the corresponding non-linear system.  The reversibility of system~\eqref{udotvdot_restricted} also guarantees, that .
%Therefore, we conclude that any equilibrium point of the original non-linear system are either a saddles or centers 
%%(for more information see  \citet{SS2018}). 
The observations so far about the equilibrium points can be summarized as follows.
\begin{lemma}
\label{lemma_restricted_equilibrium_points_position}
For system \eqref{udotvdot_restricted}, all the equilibrium points lie on the $u$-axis, and they can either be a center or a saddle. Moreover, if $\alpha<0$, there is exactly one equilibrium point $(u,0)$ with $u\in(-\infty,0)$, and if $\alpha>0$, there are three equilibrium points $(u_1,0),(u_2,0)$ and $(u_3,0)$ with $u_1\in(-\infty,0),u_2\in(0,1)$ and $u_3\in(1,\infty)$, respectively.
\end{lemma}

There is an interesting characterization of fixed configurations of vortices and equilibrium points in the $(u,v)$ phase plane as stated below.
\begin{proposition} For the restricted problem~\eqref{z0_eqn}--\eqref{z2_eqn}, the three-vortex system is in a fixed configuration if and only if the corresponding trajectory in the $(u,v)$ phase plane is an equilibrium solution.
\label{prop_fixed_config}
\begin{proof}
%\textcolor{red}{
%\begin{align*}
%r_2\text{ is a constant}\implies v(t)\equiv 0\iff v(t)=0=\dot{v}(t)\iff \dot{u}(t)=0=\dot{v}(t)\\ \iff u=\text{constant and }v=\text{constant}\implies r_2=\text{constant and }r_{12} \text{ constant} 
%\end{align*}
%}
For a fixed configuration, the lengths of the three sides of the vortex triangle should remain constant throughout the motion. Since $r_1\equiv 1$, it is enough to find what the necessary and sufficient condition is for $r_2$ and $r_{12}$ being constant functions. Using \eqref{udotvdot_restricted} one can find the derivative of $r_2^2=u^2+v^2$ in terms of $u$ and $v$ as
\begin{equation}
\label{restricted_r2sqr_dot}
\dot{r_2^2}=\frac{-\Gamma_1 v}{\pi\left((u-1)^2+v^2\right)}.
\end{equation}
Since $(u-1)^2+v^2$ is a bounded quantity (see lemma~\ref{lemma_restricted_bound} in~Sec.~\ref{subsec:Trajectories}), it follows that for a fixed configuration it is necessary that the vortex system  remains in a collinear configuration ($v=0$) for all time, i.e. $v=0$ and $\dot{v}=0$. Furthermore, since $v=0$ implies $\dot{u}=0$, the corresponding trajectory in the $(u,v)$ phase plane must therefore be an equilibrium solution. This is also a sufficient condition as $r_{12}^2=(u-1)^2+v^2$ and $r_2^2=u^2+v^2$ are  constants when $u$ and $v$ are constant functions.
\end{proof}
\end{proposition}
Let us now look at the nature of the trajectories in the $(u,v)$ phase plane.
%\begin{itemize}
%\item[a)]$k=-1$\\
%$u^3-u^2+1=0\implies u\approx -0.754878$. \\
%%u=\frac{1}{3}\left(1-\left(\frac{2}{25-3\sqrt{69}}\right)^{1/3}-\left(\frac{1}{2}\left(25-3\sqrt{69}\right)^{1/3}\right)\right)
%Hence the fixed point is at $(-0.754878,0)$. There is only one real root of the cubic polynomial as the discriminant is negative. Linear stability analysis shows that, this fixed point is an unstable saddle point.
%\item[b)]$k=1$\\
%$u^3-u^2-2u+1=0\implies u_1\approx-1.24698,\quad u_2\approx0.445042 ,\quad u_3\approx 1.80194 $.\\
%There are three fixed points on $u$-axis. $u_1$ is found out to be a linear center. $u_2$ and $u_3$ are saddle points. Since the given system is conservative (There is an equation that describes the trajectories), a linear center is a non-linear center and we can conclude that $u_1$ is an actual center. 
%\item[c)]$k\approx0$\\
%$u^3-u^2-u+1=0\implies u=\pm1$\\
%Eventhough there are two distinct roots for the cubic polynomial, in case of $k\approx0$,  we only have one fixed point situated at $u=-1$ on $u$-axis, since $u=1$ is a point of singularity and not part of the domain of the problem. Linear stability analysis shows that $u=-1$ is a degenerate point with both the eigenvalues zero. When $k$ tends to zero from right, the two saddle points starts to approach $u=1$ along $u$-axis, and exactly at $k=0$ they merge and disapear, as seen from figure \ref{figure_restricted_contour}(c). The level curves all starts to look more and more circular around the origin as expected.
%\end{itemize}
%
%
%
%
%
\subsection{Trajectories}
\label{subsec:Trajectories}

System~\eqref{udotvdot_restricted} in the Hamiltonian form reads
\begin{equation}
\dot{u}=-\frac{\partial H}{\partial v}\quad\text{and}\quad \dot{v}=\frac{\partial H}{\partial u}, 
\end{equation}
with the Hamiltonian being
\begin{equation}\label{Hamiltonian_restricted}
H(u,v)=-\frac{\omega}{2}(u^2+v^2)+\frac{\Gamma_1}{4\pi}\log\big((u-1)^2+v^2\big)+\frac{\Gamma_0}{4\pi}\log\big(u^2+v^2\big).
\end{equation}
%(\textcolor{blue}{$\dot{u}=-\partial H/\partial v$ and $\dot{v}=\partial H/\partial u$}). Integrating equation~\eqref{udot_restricted} with respect to $v$ and \eqref{vdot_restricted} with respect to $-u$ and then)
Since the Hamiltonian is a constant of the motion (see definition on page~\pageref{defn_hamiltonian}), we see that the trajectories of $\eta_2$ are precisely the level curves of~\eqref{Hamiltonian_restricted}. Note that the Hamiltonian is  a function of  $v^2$ rather than $v$, as a consequence of the reversibility of the system. The following can be observed directly from the expression of the Hamiltonian.
% In figure \ref{figure_restricted_contour}, we have shown the contours of the Hamiltonian function \eqref{Hamiltonian_restricted} for different initial conditions of $z_2$. Here, we have looked at two representative cases, to show the dependency of sign of $\Gamma_0$ and $\Gamma_1$ on the trajectory of $\eta_2$ in the $u-v$ plane. In figure \ref{figure_restricted_contour}(a) we have taken equal values of circulations ($\Gamma_0=100=\Gamma_1$) and in figure  \ref{figure_restricted_contour}(b) circulations with opposite sign ($\Gamma_0=100,\Gamma_1=-100$). All the level curves  seems to be closed in both the cases, hinting the periodicity in the variable $\eta_2$. 
\begin{lemma}
\label{lemma_restricted_bound}
 $r_2$ and $r_{12}$ are 
\begin{itemize}
\item[(i)] bounded away from zero,\quad and \quad (ii) bounded above.
\end{itemize}
\begin{proof}
Since initially $r_2$ and $r_{12}$ are assumed to be non-zero, $H$ is a finite constant given by $H(u,v)=H(u|_{t=0},v|_{t=0})$. The first part then directly follows from~\eqref{Hamiltonian_restricted}. 
%\textcolor{red}{ If one takes the limit $(u,v)\to (0,0)$ in \eqref{Hamiltonian_restricted}, the right hand side blows up, but the left hand side is finite.} 
For the second part, it suffices to show that $r_2$ is bounded above (as from triangle inequality, we have $r_{12}\leq r_2+1$) and the result will follow. To obtain a contradiction, let us assume that $r_2$ is not bounded above. This implies that there exists a sequence $\{t_n\}_{n\in\mathbb{N}}\subset \mathbb{R}$ which tends to $t^*\in \mathbb{R}\cup\{-\infty,\infty\}$ and $r_2^2(t_n)=u_n^2+v_n^2$ tends to infinity, where $u_n=u(t_n)$ and $v_n=v(t_n)$. Consider the sequence $\{H(u_n,v_n)/(u_n^2+v_n^2)\}_{n\in\mathbb{N}}$. Since $H$ is a finite constant, this sequence must converge to zero as $n$ tends to infinity. However, from expression~\eqref{Hamiltonian_restricted}, we find that the same sequence converges to $-\omega/2$, a non-zero real number. This is a contradiction and therefore $r_2$ must be bounded above.
\end{proof}
\end{lemma}

Hence from lemma~\ref{lemma_restricted_bound},  we see that in the restricted case, there are no vortex collapse situations or unbounded motions.
  
\begin{remark}(Existence of unique global smooth solutions)
\begin{figure}[!htbp]
\centering
\includegraphics[height=3in]{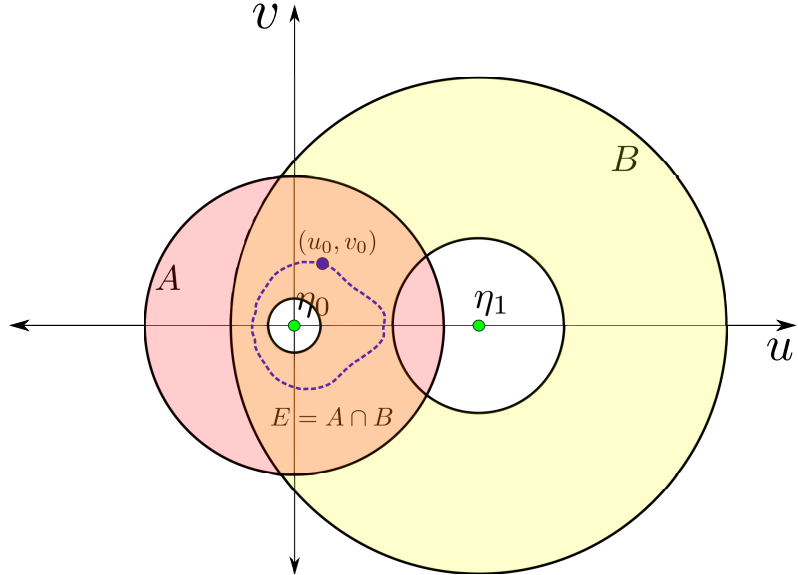}
\caption{\label{figure_smooth_trajectory}\small{Schematic showing the regions $A$ (light pink, left most annular region), $B$ (light yellow, right most annular region) and $E=A\cap B$ in $(u,v)$ phase plane. Two dots along the $u$-axis represent singularities $\eta_0$ and $\eta_1$. The dotted curve in region $E$ denotes a representative trajectory for $\eta_2(t)$ starting at some point $\eta_2|_{t=0}=(u_0,v_0)$ (
marked by a dot).}}
\end{figure}
Let $(u_0,v_0)$ be any non-singular point in the $(u,v)$ phase plane, and let $\eta_2(t)=(u(t),v(t))$ be  a solution of system~\eqref{udotvdot_restricted} with the initial condition $\eta_2|_{t=0}=(u_0,v_0)$. Local existence of such  a solution is guaranteed as $\dot{u}$ and $\dot{v}$ are continuous in a sufficiently small neighbourhood of $(u_0,v_0)$. From lemma~\ref{lemma_restricted_bound}, there exist positive constants $a_0,a_1,b_0$ and $b_1$ such that
\begin{equation}
a_0<u(t)^2+v(t)^2<b_0\quad\text{and}\quad a_1<(u(t)-1)^2+v(t)^2<b_1.
\end{equation}
Let $E=A\cap B$, see, figure~\ref{figure_smooth_trajectory}, where 
\begin{align}
A&=\{(u,v)|a_0<u^2+v^2<b_0\}\quad \mbox{and} \quad
B=\{(u,v)|a_1<(u-1)^2+v^2<b_1\}.
\end{align}
$E$ is an open set that contains the point $(u_0,v_0)$ as well as the curve $\eta_2$. Since $\dot{u}$ and $\dot{v}$ are smooth in $E$ (thus, they are continuous as well as have bounded partial derivatives in $E$), the existence and uniqueness theorem yields $\eta_2$ is the unique smooth solution of system~\eqref{udotvdot_restricted} with the non-singular initial condition $\eta_2|_{t=0}=(u_0,v_0)$ and is defined for all time.
\end{remark}

The boundedness immediately implies the following result about the trajectories.
\begin{lemma}
\label{lemma_restricted_cut_or_saddle}
%For both $t>0$ and $t<0$, a non-equilibrium trajectory of system~\eqref{udotvdot_restricted} must either cross the $u$-axis in finite time or tend to a saddle equilibrium point asymptotically.
If a trajectory of system~\eqref{udotvdot_restricted} is bounded away from the equilibrium points, then it is a closed trajectory.
\begin{proof}
%\textcolor{red}{
%\begin{equation*}
%r_2 \text{ is bounded }\implies \int_0^\infty \dot{r_2^2} \, dt<\infty\implies \lim_{t\rightarrow \infty}\dot{r_2^2}=0\implies \lim_{t\rightarrow \infty}v(t)=0
%\end{equation*}}
%WLOG, let us assume that $v(t)|_{t=0}>0$. Suppose that for $t>0$ the trajectory does not cross the $u$-axis, i.e., $v(t)>0$ for $t>0$. Using the fact that $r_2$ is bounded from both sides, and $r_{12}$ is bounded above (see lemma~\ref{lemma_restricted_bound}), from~\eqref{restricted_r2sqr_dot}, it follows that $v(t)$ must tend to zero as $t$ tends to infinity. Therefore, the point, to which the trajectory is tending to, must satisfy $v=0$ and $\dot{v}=0$, i.e., it is an equilibrium point or possibly one of the singularity points. The latter case cannot happen as seen from lemma~\ref{lemma_restricted_bound}. Hence it must be an equilibrium point. Since centers cannot have any trajectories approaching it, the equilibrium point must be a saddle point. The case $t<0$ can be shown analogously.

%Given a trajectory $\eta_2(t)=(u(t),v(t)),t\in\mathbb{R}$, consider the distance function $d(u,v)=u^2+v^2$ restricted to the compact set $C=\{(u,v)|(u,v)=\eta_2(t)\text{, for some }t\in\mathbb{R}\}$. 
Let $\eta_2(t)=(u(t),v(t)),t\in\mathbb{R}$, with $\eta_2|_{t=0}=(u_0,v_0)$, be a trajectory of system~\eqref{udotvdot_restricted} that is bounded away from the equilibrium points. Consider the distance function $d(u,v)=u^2+v^2$ restricted to the set $C=\{(u,v)|(u,v)=\eta_2(t)\text{, for some }t\in\mathbb{R}\}$. Notice that the function $d$ is continuous, 
and the set $C$ is compact (as it is bounded from lemma~\ref{lemma_restricted_bound}, and a closed set in the topological sense from being a level set of a continuous function $H(u,v)$ away from the singularities).      
According to the extreme value theorem, the continuous function $d$ attains its maximum and minimum in the compact set $C$. 
From~\eqref{restricted_r2sqr_dot}, $d$ attains maximum and minimum precisely at $v=0$, i.e. when the trajectory intersects the $u$-axis. The number of $u$-axis intersections is at least one and 
at most two, where the latter conclusion follows from the reversibility and uniqueness of trajectories. 

If there is only one intersection at the $u$-axis, the maximum and minimum must be equal and hence the distance from the origin $r_2^2(t)=d\left(u(t),v(t)\right)$ must be a constant function. Similarly, one can look at the function $r_{12}^2(t)=\left(u(t)-1\right)^2+v(t)^2$. Since
\begin{equation}
    \dot{r_{12}^2}=\frac{-\Gamma_0 v\,(u^2+v^2-1)}{\pi\left(u^2+v^2\right)},
\end{equation}
the maximum and minimum of the continuous function $r_{12}^2$ must exist either on a circle of radius one or on the $u$-axis. The first scenario cannot happen as that would imply $u^2+v^2\equiv1$ and since the trajectory has a speed $\sqrt{\dot{u}^2+\dot{v}^2}$ bounded away from zero, it will have to go through the singularity point $\eta_1$ in finite time contradicting lemma~\ref{lemma_restricted_bound}. Hence $r_{12}^2$ also attains its maximum and minimum at a point on the $u$-axis and must be a constant similar to the case of $r_2^2$. Thus, the vortex trajectory corresponds to a fixed configuration as all the inter-vortex distances are constants, which represent an equilibrium solution in the $(u,v)$ phase plane from proposition~\ref{prop_fixed_config}, a contradiction. Hence the trajectory $\eta_2(t)$ intersects the $u$-axis exactly twice. The intersection happens in finite time as from our assumption the curve has a positive non-zero minimum speed $\sqrt{\dot{u}^2+\dot{v}^2}$. Therefore, from reversibility the curve $C$ must be closed.
\end{proof}
\end{lemma}

As the points on the  $u$-axis correspond  to the collinear vortex configurations, lemma~\ref{lemma_restricted_cut_or_saddle}
essentially states that all non-fixed vortex configurations  evolve towards and away from two distinct the collinear states at any point of time. We shall now use index theory to calculate the number of saddles and centers  in the $(u,v)$ phase plane. The readers may refer the appendix for modified definitions that incorporates the presence of singularities $\eta_0$ and $\eta_1$ in system~\eqref{udotvdot_restricted} and the resulting properties. In order to find the number of saddles and centers, we shall first show that (i) the singularity points $\eta_0$ and $\eta_1$ of system~\eqref{udotvdot_restricted} have index $+1$ (in lemma~\ref{lemma_restricted_index_singularity}), and (ii) the sum of indices of all equilibrium and singularity points is one (in lemma~\ref{lemma_restricted_orbit_large_distance}).
\begin{lemma} 
The singularity points $\eta_0$ and $\eta_1$ of system~\eqref{udotvdot_restricted} have index $+1$.
\label{lemma_restricted_index_singularity}
\begin{figure}[!htbp]
\centering
\includegraphics[height=2.5in]{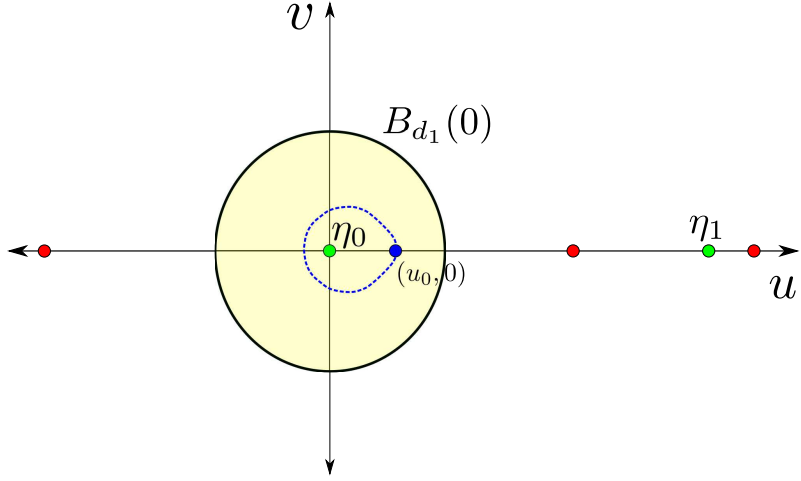}
\caption{\label{figure_index1} \small{Schematic showing the open ball $B_{d_1}(0)$ (shaded light yellow) which encloses only the singularity point $\eta_0$ and none of the equilibrium points (red filled dots) or the singularity point $\eta_1$. The blue dotted curve represents a trajectory contained in $B_{d_1}(0)$ which is initially at $(u_0,0)$ (blue filled dot).} }
\end{figure}
\begin{proof}

We shall prove this only for the case of $\eta_0$ as similar lines of arguments can be given for the case of $\eta_1$. As any closed trajectory has index $+1$ (see theorem~\ref{theorem_app:closed_trajectory} of Appendix~\ref{appendix:index}), it suffices to show that there exists a closed trajectory of system~\eqref{udotvdot_restricted} that encloses $\eta_0$, but none of the equilibrium points or the singularity point $\eta_1$. Let us define
\begin{equation}
\label{eq_singularity_distance}
d_1=\frac{1}{2}\,\min \left\{\sqrt{u^2+v^2}\, \Big\vert\, (u,v) \text{ is an equilibrium point of system}~\eqref{udotvdot_restricted}\right\}.
\end{equation} 
%(Recall that equilibrium points lie on the $u$-axis, i.e., in equation~\ref{eq_singularity_distance}, $v\equiv 0$).
By the definition of $d_1$, the open ball, $B_{d_1}(0)=\{(u,v)\in\mathbb{R}^2|u^2+v^2< d_1^2\}$ does not contain any of the equilibrium points. Since $p(-1)=\alpha,\, p(0)=1$ and $p(1)=-\alpha$, the polynomial $p$ always has a real root in $|u|<1$ for any $\alpha\neq0$. Hence it follows from the definition of $d_1$ that  $d_1<1/2$  and $\eta_1\notin B_{d_1}(0)$, as illustrated in figure~\ref{figure_index1}. 

We shall first show that there exists a trajectory that is fully contained in $B_{d_1}(0)$. From~\eqref{restricted_r2sqr_dot}, it follows that the maximum and minimum distances of a trajectory from the origin is attained for $v=0$, i.e., when a trajectory intersects the $u$-axis. Moreover, it is a maximum if the sign of the  double derivative of $r_2^2$ is negative, i.e., if  $\Gamma_1\dot{v}>0$ at the point where the trajectory intersects the $u$-axis. Hence, it is enough to find a point on the $u$-axis which lies inside $B_{d_1}(0)$ and satisfies $\Gamma_1\dot{v}>0$, since the unique trajectory passing through such a point will be contained in $B_{d_1}(0)$ for all time.
 % Let us consider all  the trajectories that start at the $u$-axis, i.e., $\eta_2(t=0)=(u_0,0)$ for some $u_0\in \mathbb{R}$. For these trajectories, from~\eqref{udotvdot_restricted} we have
For a point $(u_0,0)$ on the $u$-axis,  from~\eqref{udotvdot_restricted} we have
\begin{equation}
\label{restricted_vdot_on_u_axis}
\dot{v}|_{(u,v)=(u_0,0)}=\frac{\Gamma_0}{2\pi}\left(-u_0+\frac{1}{u_0}+\frac{\alpha}{u_0-1}\right)
=\frac{\Gamma_0}{2\pi}\frac{p(u_0)}{u_0(1-u_0)}.
\end{equation}
%Since $p(0)=1>0$, and from the continuity of the cubic polynomial $p$, we know that in a sufficiently small neighbourhood of zero, $p(u_0)$ is positive.
If we choose $u_0$ from a sufficiently small neighbourhood of zero, then $p(u_0)>0$, because of the continuity of the cubic polynomial $p$ and the fact that $p(0)=1>0$. Depending on the sign of $\Gamma_1$ and $\Gamma_0$, one could then choose a negative or positive $u_0$ from this neighbourhood of the origin that satisfies $\Gamma_1\dot{v}>0$.    
Thus, it is always possible to find a point $(u_0,0)\in B_{d_1}(0)$ such that the unique trajectory that starts at $(u_0,0)$ satisfies $u^2+v^2<u_0^2<d_1^2$, i.e., this trajectory will be contained in $B_{d_1}(0)$ for all time. Since this trajectory is bounded away from the equilibrium points by construction~\eqref{eq_singularity_distance}, it must be a closed trajectory from lemma~\ref{lemma_restricted_cut_or_saddle}. Finally, we note that since all closed trajectories must enclose at least one singularity or an equilibrium point, the trajectory we constructed  encloses the origin.
\end{proof}
\end{lemma}

\begin{remark}
Let $C$ be the closed trajectory that we have constructed while giving the proof of lemma~\ref{lemma_restricted_index_singularity}. Consider any point $(u_0,v_0)$ in the interior of $C$.  The unique trajectory that starts at $(u_0,v_0)$  must be contained in the interior of  $C$, as no two trajectories can ever intersect. Therefore the arguments used in lemma~\ref{lemma_restricted_index_singularity} can be repeated to conclude  that all trajectories in the interior of $C$ are closed. Hence both the singularity points $\eta_0$ and $\eta_1$ have a region of closed trajectories surrounding them.
\end{remark}
 One can use similar arguments to show that the sum of indices of  equilibrium and singularity points is one.
\begin{lemma}The sum of indices of all equilibrium and singularity points of system~\eqref{udotvdot_restricted} is +1.
\label{lemma_restricted_orbit_large_distance}
\begin{figure}[!htbp]
\centering
\includegraphics[height=3in]{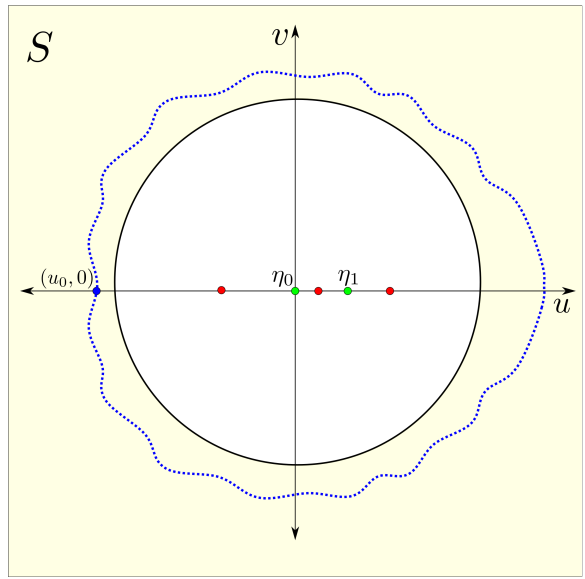}
\caption{\label{figure_index2} \small{Schematic showing the region $S$ (light yellow shaded) which does not contain any of the singularity points $\eta_0$ and $\eta_1$ (green dots) or equilibrium points (red dots). The blue dotted curve represents a trajectory contained in $S$ that starts at the point $(u_0,0)$ (blue filled dot on the curve). }}
\end{figure}
\begin{proof}
It is enough to show that, there exists a closed trajectory that encloses the singularity points $\eta_0$ and $\eta_1$ as well as the equilibrium points of system~\eqref{udotvdot_restricted} (follows from theorem~\ref{theorem_app:sum_index} and \ref{theorem_app:closed_trajectory} of Appendix~\ref{appendix:index}).
We define
\begin{equation}
d_2=2\,\max \left\{\sqrt{u^2+v^2}\, \Big\vert\, (u,v) \text{ is an equilibrium point of system}~\eqref{udotvdot_restricted}\right\}.
\end{equation}
The set $S=\{(u,v)\in\mathbb{R}^2|u^2+v^2> d_2^2\}$ contains none of the equilibrium  or the singularity points. We shall proceed just as in the proof of lemma~\ref{lemma_restricted_index_singularity}, i.e., we will try first to establish that there exists a trajectory that is fully contained in $S$. This is accomplished by finding a point $(u_0,0)$ in $S$,  satisfying $\Gamma_1\dot{v}<0$. The unique trajectory that passes through such a point $(u_0,0)$ has its minimum distance from the origin  attained at $(u_0,0)$, and hence the trajectory will satisfy $u^2+v^2\geq u_0^2>d_2^2$, as required, see the dotted trajectory in figure~\ref{figure_index2}. 
Note that for sufficiently large $|u_0|$, from expression~\eqref{restricted_vdot_on_u_axis} we have $\Gamma_1\dot{v}\approx-\Gamma_0\Gamma_1 u_0/2\pi$. This can be made negative by appropriately taking $u_0$ positive or negative. Consequently, one can always find a trajectory  bounded away from the equilibrium points. Therefore, from lemma~\ref{lemma_restricted_cut_or_saddle}, such a trajectory must be a closed trajectory. Since a closed trajectory must enclose at least one equilibrium point or a singularity point, the only way it can happen is when the  trajectory we constructed encloses all of them. 
\end{proof}
\end{lemma}

\begin{remark}
Let $C$ be the closed trajectory that we have constructed in the proof of lemma~\ref{lemma_restricted_orbit_large_distance}. Consider any point $(u_0,v_0)$ in the exterior of $C$. The unique trajectory that passes through $(u_0,v_0)$ is bounded by $C$ and hence bounded away from the equilibrium points. Therefore, one can repeat the arguments from lemma~\ref{lemma_restricted_orbit_large_distance}  to conclude that all trajectories in the exterior of $C$ are closed and encloses all the equilibrium and singularity points.
\end{remark}

A saddle has an index $-1$ whereas a center has an index $+1$~\cite{SS2018}. Let $m$ be the number of centers and $n$ be the number of saddles. From lemma~\ref{lemma_restricted_orbit_large_distance}, we know that there exists a closed orbit which encloses all the equilibrium points and singularities. Since any closed orbit in the phase plane must enclose points whose indices sum to $+1$, we have 
\begin{equation}
\label{restricted_index_sum}
m\times \underbrace{1}_{\text{center}}+ \quad n\times\underbrace{-1}_{\text{saddle}}+\underbrace{1}_{\eta_0}+\underbrace{1}_{\eta_1}=\underbrace{1}_{\text{total}}\implies m-n=-1.
\end{equation} 
For $\alpha<0$ there is only one equilibrium point and hence $m+n=1$. Together with~\eqref{restricted_index_sum}, this linear system yields a unique integer solution given by $m=0$ and $n=1$. Therefore, for $\alpha<0$, the only equilibrium point is a saddle point. Similarly, for $\alpha>0$ there are three equilibrium points, i.e., $m+n=3$. The linear system so obtained together with~\eqref{restricted_index_sum} gives $m=1$ and $n=2$ as the unique solution. Therefore, for $\alpha>0$, there are  two saddles and one center. 

Let us now focus our attention to the case of $\alpha>0$.  We are interested in finding the relative positions of the center and saddle equilibrium points on the $u$-axis. This can be achieved  by analyzing the Hamiltonian values of the three equilibrium solutions, as explained below. 

Since system~\eqref{udotvdot_restricted} is invariant under the transformation $(t,\Gamma_0,\Gamma_1)\to(-t,-\Gamma_0,-\Gamma_1)$, one can assume WLOG that $\Gamma_0>0$. As we are now considering the case  $\alpha>0$, this would imply $\Gamma_1>0$. Note that, (i) the Hamiltonian, $H$, is continuous everywhere except at the singularity points $(0,0)$ and $(1,0)$, where the function tends to negative infinity. Hence $H$ is an upper semi-continuous function.  (ii) The critical points of the Hamiltonian $H$ are precisely the equilibrium points of system~\eqref{udotvdot_restricted}.  From lemma~\ref{lemma_restricted_orbit_large_distance}, there exists a closed trajectory $C$ which encloses all three equilibrium points. Since the trajectories in the exterior of  $C$ are also closed (see, remark after lemma~\ref{lemma_restricted_orbit_large_distance}) and $H(u,v)$ tends to negative infinity as $u^2+v^2$ tends to infinity, one can always find a closed trajectory $C'$ with  $H(C')$  smaller than any of the Hamiltonian values corresponding to the equilibrium points. Let the bounded interior of $C'$ be denoted by $\text{int}(C')$. The set $D= \text{int}(C')\cup C'$ is a compact set. Since an upper semi-continuous function attains its supremum in a compact set, the Hamiltonian function $H(u,v)$ has a maximum in $D$.  The maximum cannot occur on $C'$ from construction. Therefore it must be in the interior of $D$ and  a local maximum  of the Hamiltonian function $H$.   For a Hamiltonian system with isolated equilibrium points, local minima and maxima are centers~\cite{SS2018}. Consequently, among the three equilibrium points, the center is precisely the one with the maximum Hamiltonian value.

Since all the equilibrium points have their $v$ component zero, it is enough to look at the real-valued function  
\begin{equation}
f(u,\alpha)=\log(u^2)-u^2+\alpha \log\left((u-1)^2\right)
\end{equation}
to compare the Hamiltonian values of the three equilibrium points for different $\alpha$ values. %Note that the actual Hamiltonian is just a constant multiple of the above function restricted to $u$-axis. 
For $\alpha>0$, one can also define the functions 
\begin{align}
\left.
\begin{aligned}
u_1(\alpha)&=\text{root of the polynomial }p, \text{that lies in }(-\infty,0),\\
u_2(\alpha)&=\text{root of the polynomial }p, \text{that lies in }(0,1),\\
u_3(\alpha)&=\text{root of the polynomial }p, \text{that lies in }(1,\infty).
\end{aligned}
\right\}
\end{align}
We would like to explore the behaviour of the functions $g_j(\alpha)=f(u_j(\alpha),\alpha)$ for $j\in\{1,2,3\}$ and identify the center. It can be easily verified that 
\begin{equation}
\lim_{\alpha\rightarrow0}g_1(\alpha)=\lim_{\alpha\rightarrow0}g_2(\alpha)=\lim_{\alpha\rightarrow0}g_3(\alpha)=-1.
\end{equation} Differentiating $g_j(\alpha)$ with respect to $\alpha$ we get,
\begin{align}
\left.
\begin{aligned}
\frac{dg_j}{d\alpha}&=\frac{\partial }{\partial \alpha}(f(u_j,\alpha))+\frac{\partial }{\partial u_j}(f(u_j,\alpha))\times \frac{du_j}{d\alpha},\\
&=\log\left((u_j-1)^2\right)-\frac{2\,p(u_j)}{u_j(u_j-1)}\times \frac{du_j}{d\alpha},\\
&=\log\left((u_j-1)^2\right)-0\times \frac{du_j}{d\alpha},\\
&=\log\left((u_j-1)^2\right).
\end{aligned}
\right\}
\end{align}
Since $u_1(\alpha)\in(-\infty,0)$, $d(g_1)/d\alpha$ is strictly positive. This further implies that  $g_1$ is a monotonically increasing function of $\alpha$. On the other hand, as $u_2\in(0,1)$, the function $g_2$ is monotonically decreasing. Hence $g_2(\alpha)$ can never be larger than $g_1(\alpha)$ and hence $u_2$ cannot be a center (recall that the center has the maximum Hamiltonian value).  Therefore, it must be a saddle equilibrium point for all $\alpha>0$. 

For $g_3(\alpha)$,  $p(2)=3 - 2 \alpha$ and from intermediate value theorem one can conclude
\begin{equation}
u_3(\alpha)\in \begin{cases}(1,2),& \text{if }\alpha<3/2,\\
\{2\},& \text{if }\alpha=3/2,\\
(2,\infty),& \text{if } \alpha>3/2.
\end{cases}
\end{equation}
Therefore, the function $g_3(\alpha)$ is monotonically decreasing ($dg_3/d\alpha<0$) for $\alpha<3/2$ and  strictly increasing ($dg_3/d\alpha>0$) for $\alpha>3/2$. Hence if $\alpha\in (0,3/2]$,  $g_1(\alpha)$ is larger than both $g_2(\alpha)$ and $g_3(\alpha)$. So, $u_1$ must be a center at least when $\alpha \in(0,3/2]$.  Suppose  $u_3$ is a center for some $\alpha_0>3/2$. This means that $g_3(\alpha_0)>g_1(\alpha_0)$. Hence from the intermediate value theorem, there exists $\alpha\in(3/2,\alpha_0)$ such  that $g_1(\alpha)=g_3(\alpha)$, which further implies that there are two center equilibrium points. This is a contradiction to the fact that there is exactly one center and two saddles for $\alpha>0$. Hence $g_1(\alpha)$ stays larger than both $g_2(\alpha)$ and $g_3(\alpha)$ for all $\alpha>0$. In other words $u_1$ is always a center and the remaining two are always saddles. This is summarized in the following lemma.
\begin{lemma} 
\label{lemma_restricted_saddles_centers_numbers}
For $\alpha<0$, the only equilibrium point of system~\eqref{udotvdot_restricted} is a saddle. For $\alpha>0$, there are two saddles and one center with $u_1\in(-\infty,0) $ always being the center.
\end{lemma}
We could also use the above constructions to show that heteroclinic orbits are absent in the $(u,v)$ phase plane.
\begin{lemma}
\label{lemma_restricted_heteroclinic_orbits}
For the restricted problem~\eqref{z0_eqn}--\eqref{z2_eqn}, there are no heteroclinic orbits in the $(u,v)$ phase plane. 
\begin{figure}[!htbp]
\centering
\includegraphics[height=2.7in]{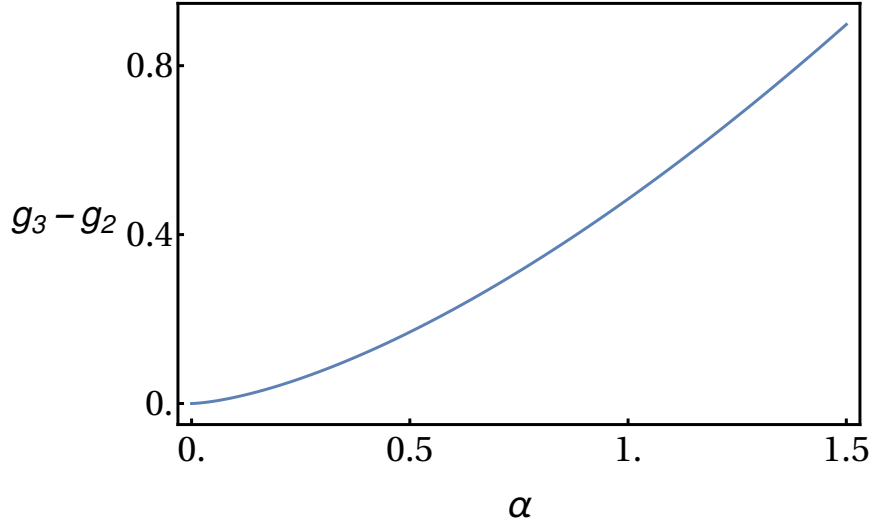}
\caption{Variation of  $g_3-g_2$ with $\alpha$.}
\label{figure_restricted_heteroclinic_orbits}
\end{figure}
\begin{proof}
%
%Since there is only one equilibrium point for the case of $\alpha<0$, we only need to consider the case of $\alpha>0$. It is enough to show that $g_2(\alpha)< g_3(\alpha)$ for  $\alpha\in(0,3/2)$, since $g_2$ is monotonically decreasing for all $\alpha>0 $ and $g_3$ is monotonically increasing for $\alpha>3/2$. This is found out to be the case (see figure~\ref{figure_restricted_heteroclinic_orbits}) numerically. Hence for the restricted problem~\eqref{z0_eqn}--\eqref{z2_eqn}, there are no heteroclinic orbits in the $(u,v)$ phase plane.
A heteroclinic orbit is a trajectory that connects two saddle equilibrium points asymptotically. Since all trajectories are level curves of Hamiltonian, the existence of a heteroclinic orbit necessitates the existence of two saddle points with the same Hamiltonian value. As for $\alpha<0$, there is only one saddle point in the phase plane (see, lemma~\ref{lemma_restricted_saddles_centers_numbers}), we need to consider only the case $\alpha>0$, i.e., it suffices to show that $g_2(\alpha)\neq g_3(\alpha)$ for any $\alpha>0$. Since $g_2$ is monotonically decreasing for all $\alpha>0 $ and $g_3$ is monotonically increasing for $\alpha>3/2$, it is enough to show that the two curves do not intersect for $\alpha\in (0,3/2]$. This is verified numerically in figure~\ref{figure_restricted_heteroclinic_orbits}. Hence for the restricted problem~\eqref{z0_eqn}--\eqref{z2_eqn}, there are no heteroclinic orbits in the $(u,v)$ phase plane.
\end{proof}
\end{lemma}
From the above lemma, we see that there cannot be any heteroclinic orbits in the $(u,v)$ phase plane and therefore, any non-equilibrium trajectory must cross the $u$-axis at least once for some $t\in \mathbb{R}$.
\begin{theorem}
\label{therorem_trajectories_uv}
For the restricted problem~\eqref{z0_eqn}--\eqref{z2_eqn}, any non-equilibrium trajectory in the $(u,v)$ phase plane is either  a closed trajectory or a homoclinic orbit.
\begin{proof}
Directly follows from lemmas~\ref{lemma_restricted_cut_or_saddle} and~\ref{lemma_restricted_heteroclinic_orbits}.\end{proof}
\end{theorem}

The above theorem yields two important insights about the vortex motion in the restricted three vortex problem~\eqref{z0_eqn}--\eqref{z2_eqn}. 
%Firstly, for all non-fixed configurations of vortices (i.e., a non-equilibrium trajectory in the $(u,v)$ phase plane, see proposition~\ref{prop_fixed_config}), the vortex $\mathcal{V}_2$ either moves in such a way that its distances from the vortices $\mathcal{V}_0$ and $\mathcal{V}_1$ are periodic (and bounded from lemma~\ref{lemma_restricted_bound}) in time (i.e., a closed orbit in the $(u,v)$ phase plane) or it is bounded but aperiodic and moves asymptotically to a collinear fixed configuration (i.e., some portion of a homoclinic orbit in the $(u,v)$ phase plane). Secondly, to study the restricted vortex problem, it is enough to look at the initial conditions that are  collinear. 
Firstly, according to theorem~\ref{therorem_trajectories_uv}, there are exactly three types of trajectories possible in the $(u,v)$ phase plane. It is either (i) an equilibrium point, (ii) a closed trajectory or (iii) a homoclinic orbit (a trajectory that tends to a saddle equilibrium point asymptotically for both $t>0$ and $t<0$). This classification physically translates to three types of vortex motions: (i) a fixed configuration of vortices (see, proposition~\ref{prop_fixed_config}), (ii) the vortex $\mathcal{V}_2$  moves in such a way that its distances from the vortices $\mathcal{V}_0$ and $\mathcal{V}_1$ ($r_2$ and $r_{12}$) are periodic in time, and (iii) $r_2$ and $r_{12}$ are aperiodic, with  vortex $\mathcal{V}_2$  asymptotically approaching a fixed configuration. 
Secondly, all three types of $(u,v)$ phase plane trajectories intersect the $u$-axis, implying that all vortex motion must go through a collinear configuration. Hence, while studying restricted three vortex problem, it is sufficient to consider initial vortex configurations that are collinear.

Let us make an interesting observation regarding numerically computing the vortex trajectories. For the sake of simplicity, we shall assume that initial conditions are all taken as collinear vortex configurations (i.e., points on the $u$-axis). Since except for a finite number of cases all initial conditions lead to a closed orbit in the $(u,v)$ phase plane, directly doing a numerical simulation of system \eqref{z0_eqn}--\eqref{z2_eqn} may potentially mislead us into thinking that inter vortex distances are always periodic. The bias is more likely when the aperiodic trajectories have initial conditions at irrational points. Since in computers numbers are represented in the floating-point format that essentially gives rationals, none of the numerical solutions generated with random initial conditions would accurately represent aperiodic solutions.  Thus, one needs to use a high working precision along with the exact values of initial conditions to even approximate  aperiodic vortex trajectories through numerical solver, such as the Euler or the  Runge--Kutta methods. 

\begin{remark}{(Period of a closed $(u,v)$ phase plane orbit)} It is enough to consider a single variable period function $u_0\mapsto T(u_0)$, defined as the period of the closed orbit that is initially at $(u_0,0)$.  Note that there is at least one saddle equilibrium point in the phase plane (see lemma~\ref{lemma_restricted_saddles_centers_numbers}) so that  $T(u_0)$ tends to infinity as $u_0$ tends to this saddle point. Also, as $u_0$ approaches $\eta_0$, the trajectories speed up (because of the singularity at zero) and become smaller (see remark after lemma~\ref{lemma_restricted_index_singularity}), indicating  that $T(u_0)$ tends to zero as $u_0$ tends to zero.  Hence by continuity, the range of $T$ is $(0,\infty)$.
\end{remark}
\section{Examples}
%-------------------------

\begin{figure}[!htbp]
\centering
\subfigure[]{
\includegraphics[height=2.5in]{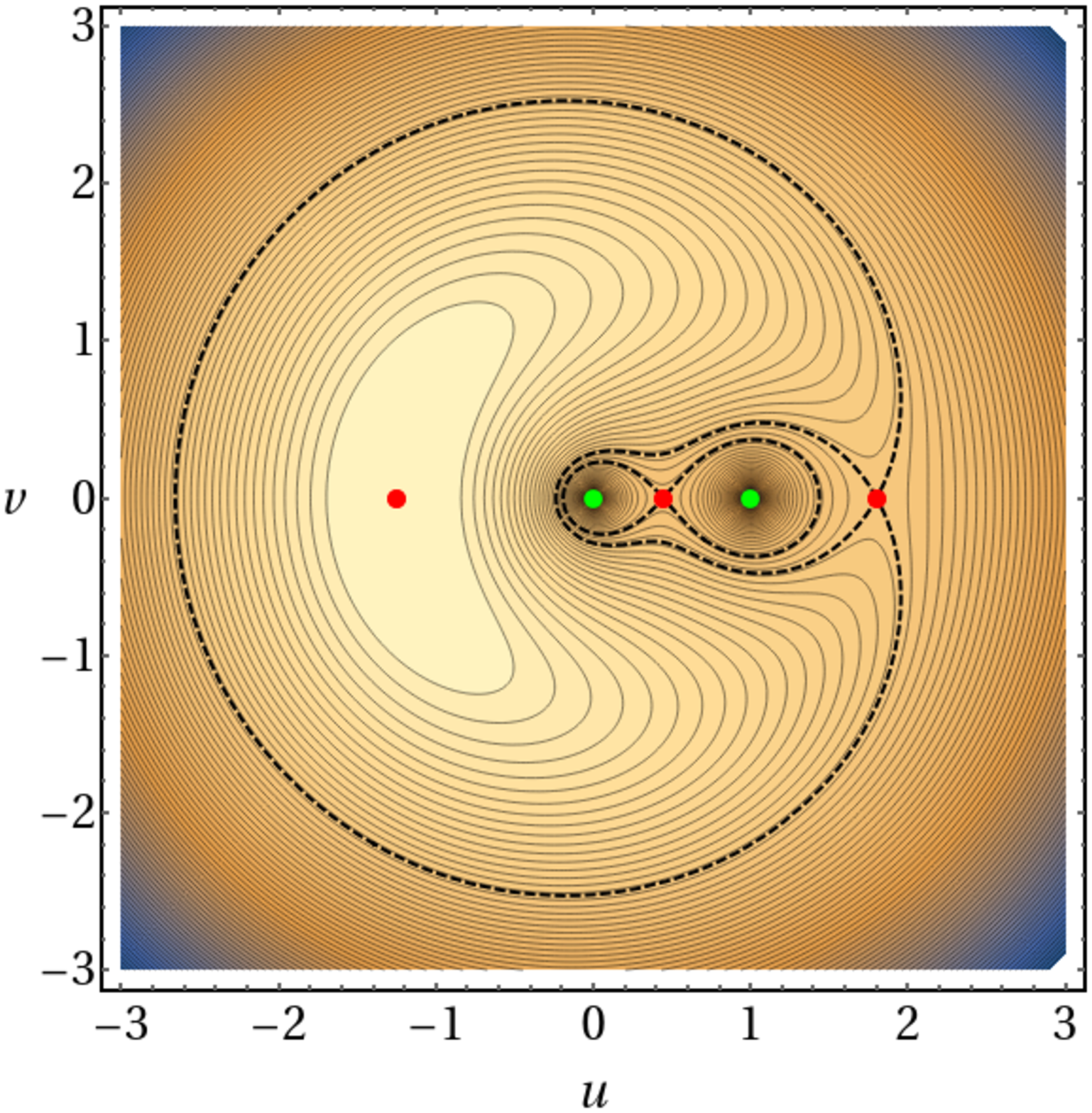}}\qquad\qquad\qquad
\label{figure_restricted_equal}
\subfigure[]{\includegraphics[height=2.5in]{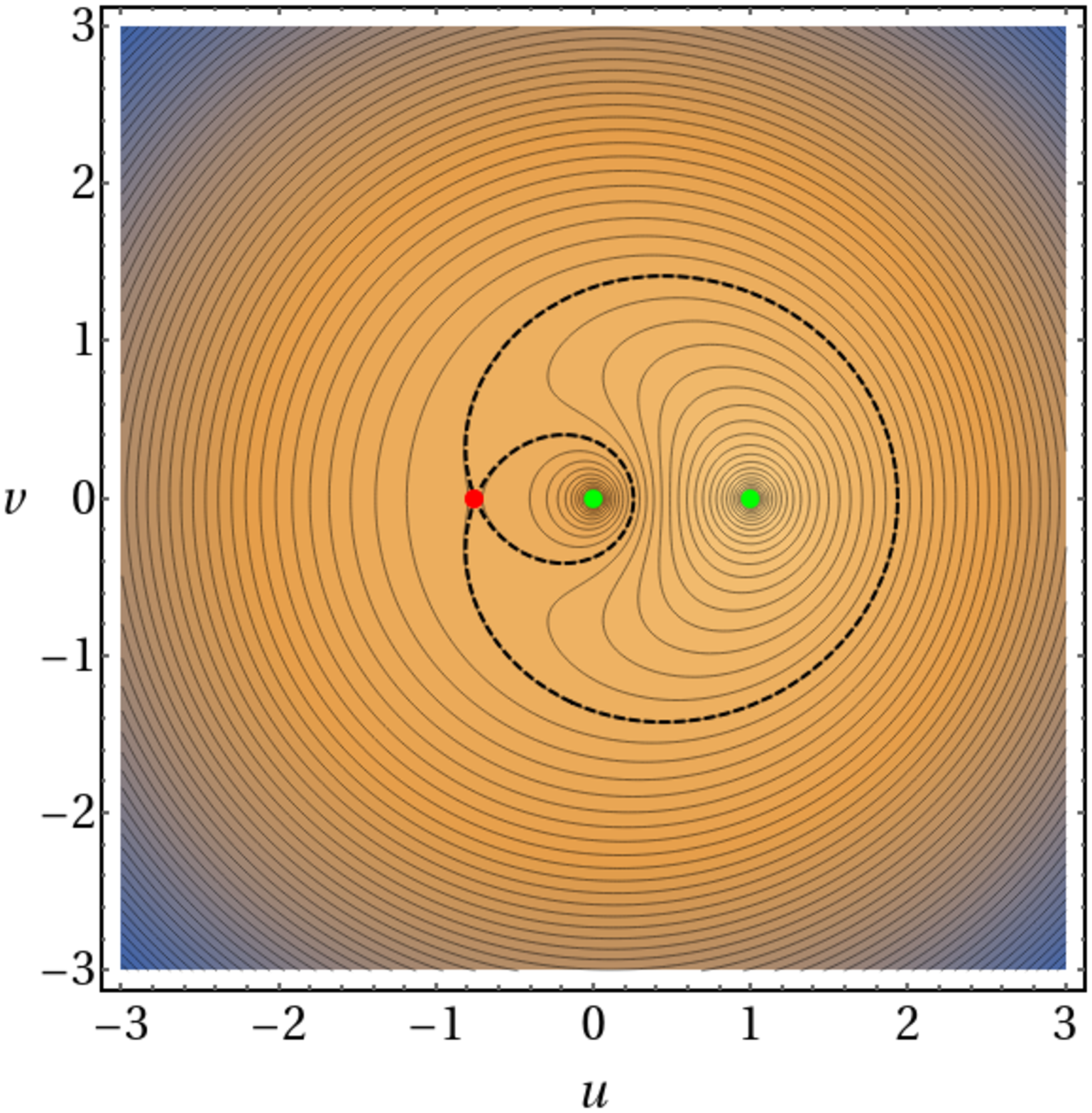}}
\label{figure_restricted_opposite}
%\subfigure[]{
%\includegraphics[height=2.5in]{figures/3.eps}}\qquad
%\label{figure_restricted_notequal1}
%\subfigure[]{\includegraphics[height=2.5in]{figures/4.eps}}
%\label{figure_restricted_notequal2}
\caption{\label{figure_restricted_contour}\small{Level curves of the Hamiltonian function \eqref{Hamiltonian_restricted} for (a) $\alpha=1$, and (b) $\alpha=-1$.
%(c) $\Gamma_1=1$ (d) $\Gamma_1=-1$. 
Positions of singularity points $\eta_0$  and $\eta_1$ are marked with green dots, while the equilibrium points are marked with red dots. The black dashed lines indicate the homoclinic orbits.}}
\end{figure}

We shall now consider two representative cases for the  positive and negative values of $\alpha=\Gamma_1/\Gamma_0$. A contour plot of the Hamiltonian~\eqref{Hamiltonian_restricted} for $\alpha=\pm 1$ yields the $(u,v)$ phase plane trajectories as illustrated in figure~\ref{figure_restricted_contour}. One may recall that these are the trajectories of vortex $\mathcal{V}_2$ in the rotating frame of reference for different initial conditions. The equilibrium points are marked with red dots, whereas  the positions of $\mathcal{V}_0$ and $\mathcal{V}_1$ in the rotating frame of reference  are marked with green dots. The black dashed lines are the homoclinic orbits.
%\subsection{\texorpdfstring{$\alpha=1$}{Lg}}
The exact coordinates of the equilibrium points are found by solving the cubic polynomial~\eqref{u_fixedpoint_cubic_polynomial}. For $\alpha=1$, the real roots are located at $u=u_1\approx -1.24698$, $u=u_2\approx0.445042$ and $u=u_3\approx 1.80194$ respectively, as indicated by three red dots in figure~\ref{figure_restricted_contour}(a). There is only one real root $u\approx -0.754878$ for $\alpha=-1$, as marked by one red dot in figure~\ref{figure_restricted_contour}(b). As expected, for $\alpha=1$ there are three equilibrium points (see lemma~\ref{lemma_restricted_equilibrium_points_position} and \ref{lemma_restricted_saddles_centers_numbers}) with $u_1$ being a center and the rest being the saddle points, whereas for the case of $\alpha=-1$, there exits (the only) one saddle equilibrium point at location $u\approx -0.754878$. In agreement with theorem~\ref{therorem_trajectories_uv}, we see that except for the black dashed lines (homoclinic orbits) all other non-equilibrium trajectories are closed orbits. We shall now give examples for each of the  three different types of  vortex motion.

%--------------------------------------
\subsection{Fixed configuration case}
%--------------------------------------
\begin{figure}[htbp]
\centering
\subfigure[]{
\includegraphics[height=2.4in]{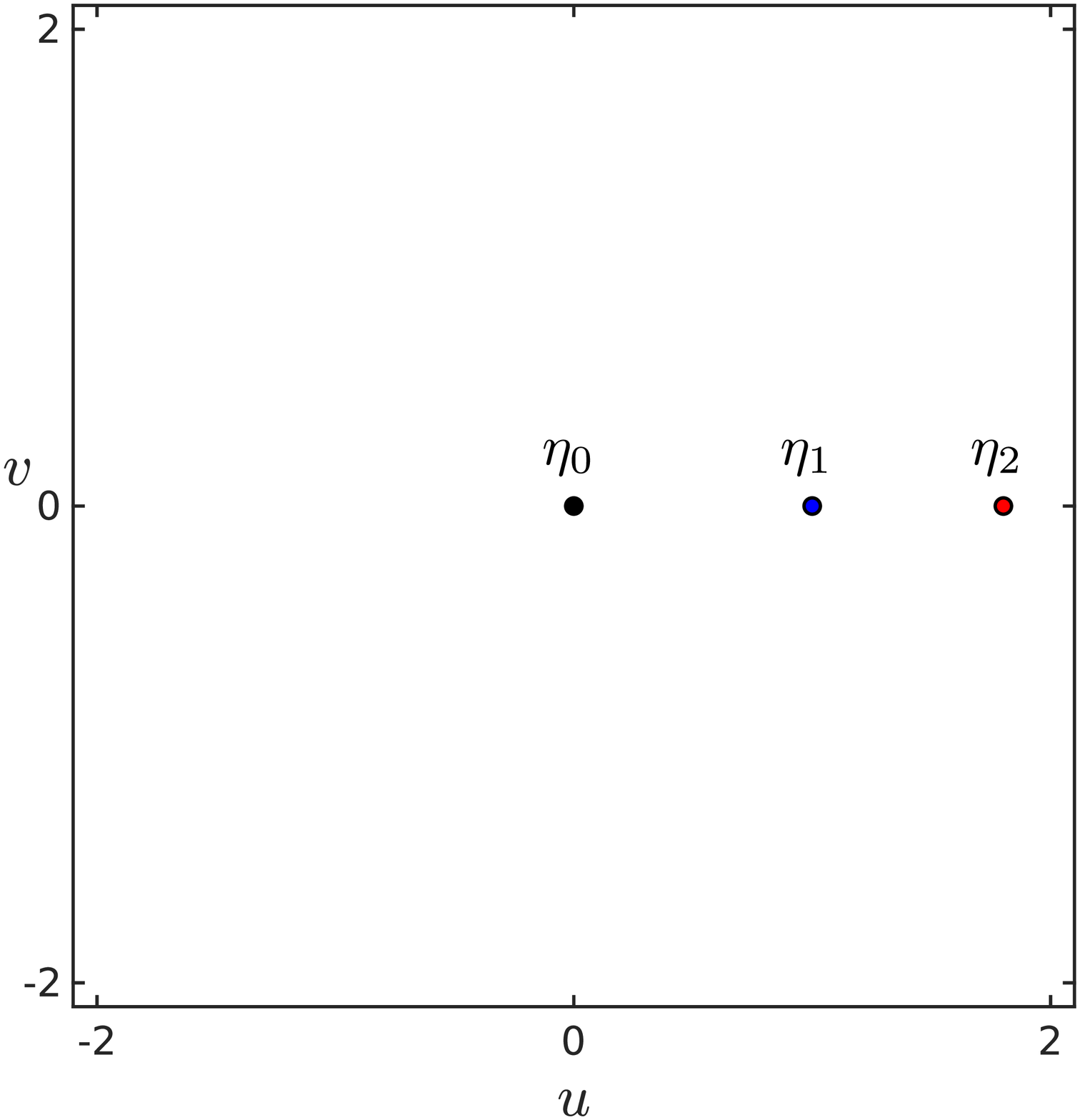}}\qquad\qquad\qquad
\label{figure_fixed_conf_uv}
\subfigure[]{\includegraphics[height=2.4in]{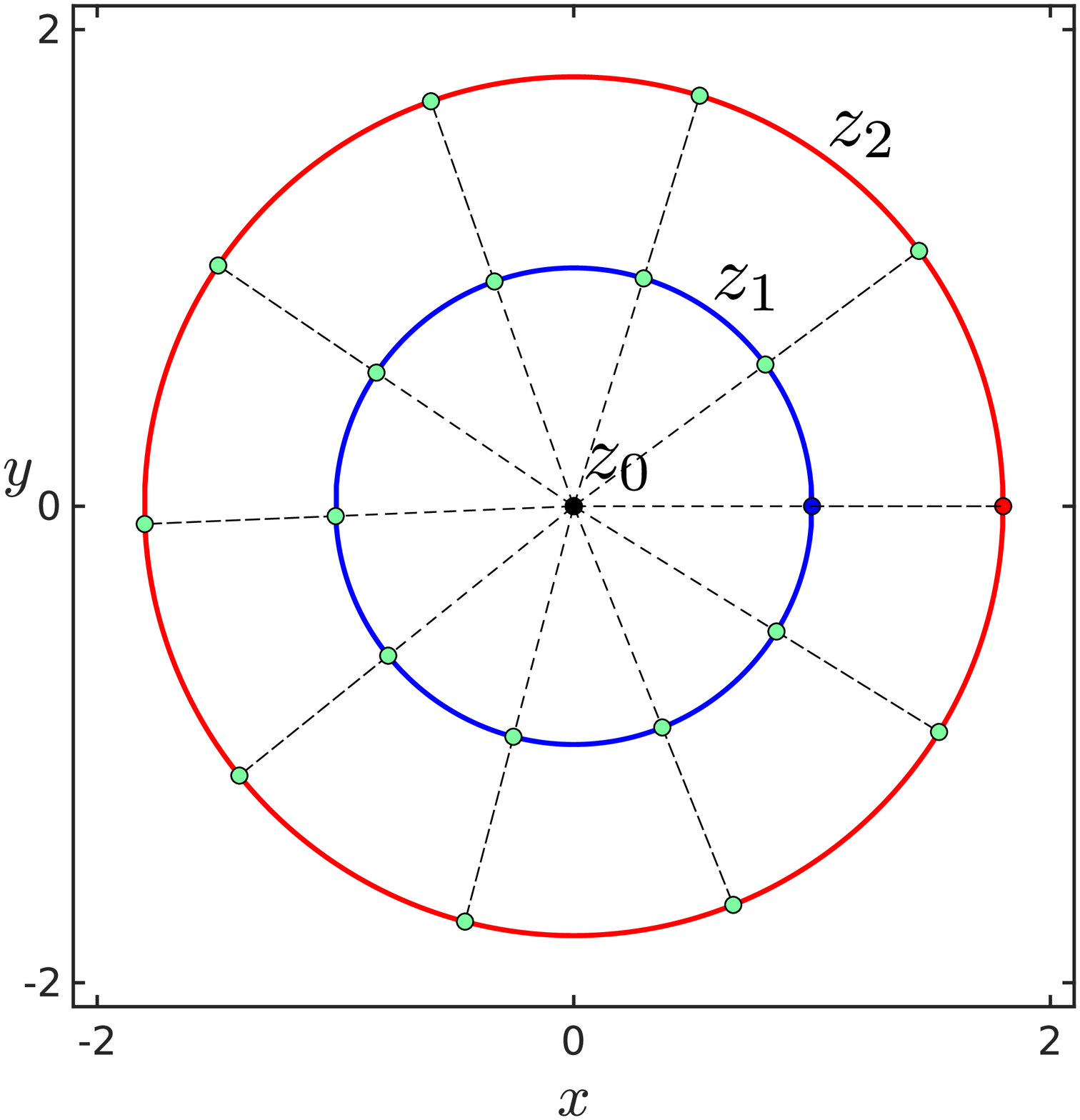}}
\label{figure_fixed_conf_xy}
\caption{\label{figure_fixed_conf}\small{An example of a fixed configuration of vortices ($\Gamma_0=\Gamma_1=10,\Gamma_2=0,z_1|_{t=0}=1,z_2|_{t=0}=\eta_2|_{t=0}=u_3\approx 1.80194, z_0=0$). The  positions of the three vortices are shown with respect to the (a) rotating and (b) stationary frame of references. The vortices $\mathcal{V}_1$ (blue trajectory) and $\mathcal{V}_2$ (red trajectory) remain collinear with the fixed vortex $\mathcal{V}_0$ at any point of time. The collinearity is illustrated by marking their positions (green) at different times and joining them (dashed lines). Black, blue  and red dots denote the initial positions $\eta_0$, $\eta_1$ and  $\eta_2$ in panel (a) as well as vortices $\mathcal{V}_0$, $\mathcal{V}_1$ and $\mathcal{V}_2$ in panel (b), respectively. }}
\end{figure}

As per proposition~\ref{prop_fixed_config}, any equilibrium solution in the $(u,v)$ phase plane corresponds to a fixed configuration and vice versa. Let us consider one such equilibrium initial condition, say, $\Gamma_0=10, \Gamma_1=10$ (i.e., $\alpha=1$) and $\eta_2|_{t=0}=u_3\approx 1.80194$. Figure~\ref{figure_fixed_conf}(a) illustrates singularity points and an equilibrium solution $\eta_2$ in $(u,v)$ (rotating) phase plane. We have already assumed $z_1|_{t=0}=1$, so that  vortex $\mathcal{V}_1$ moves in a circular path with constant angular velocity $\omega=\Gamma_0/2\pi$. As the initial condition corresponds to an equilibrium solution in the variable $\eta_2$, see figure~\ref{figure_fixed_conf}(a), we have $z_2=\eta_2\,z_1=u_3\,e^{i\omega t}$. Therefore, one would expect $\mathcal{V}_2$ to also move in a circular orbit around $\mathcal{V}_0$  with  a constant  angular velocity same as that of $\mathcal{V}_1$ but  with a different radius $u_3$. Since the vortices are collinear initially, they stay in a collinear configuration for all time. This is shown in figure~\ref{figure_fixed_conf}(b), where we have marked the positions of the three vortices at equal intervals of time $(\Delta t=0.4)$. Evidently, all three vortices lie on a straight line. 

%-----------------------------------------
\subsection{Aperiodic non-fixed configuration case}
%-----------------------------------------

\begin{figure}[htbp]
\centering
\subfigure[]{
\includegraphics[height=2.6in]{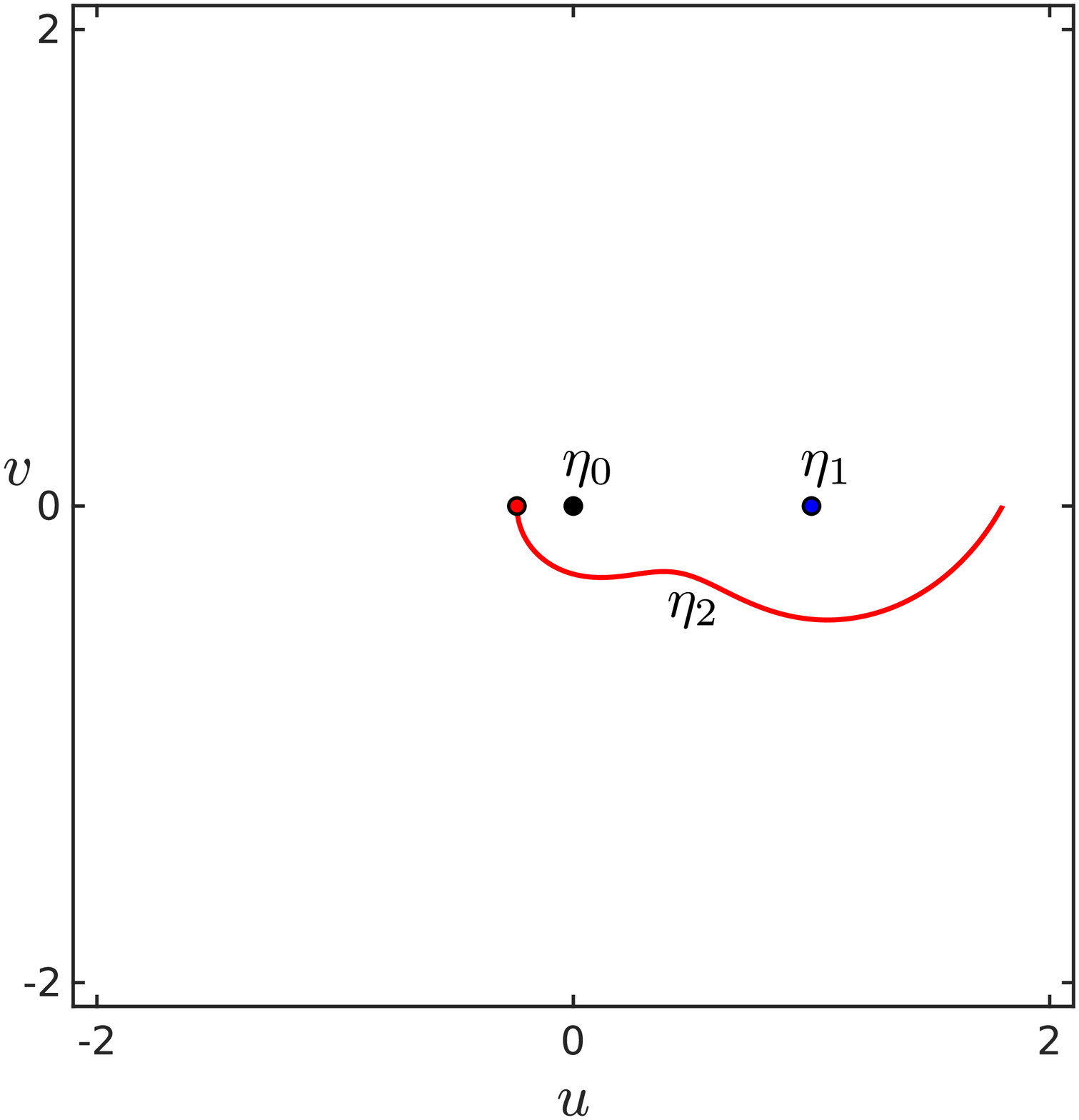}}\qquad\qquad\qquad
\label{figure_aperiodic_uv}
\subfigure[]{\includegraphics[height=2.6in]{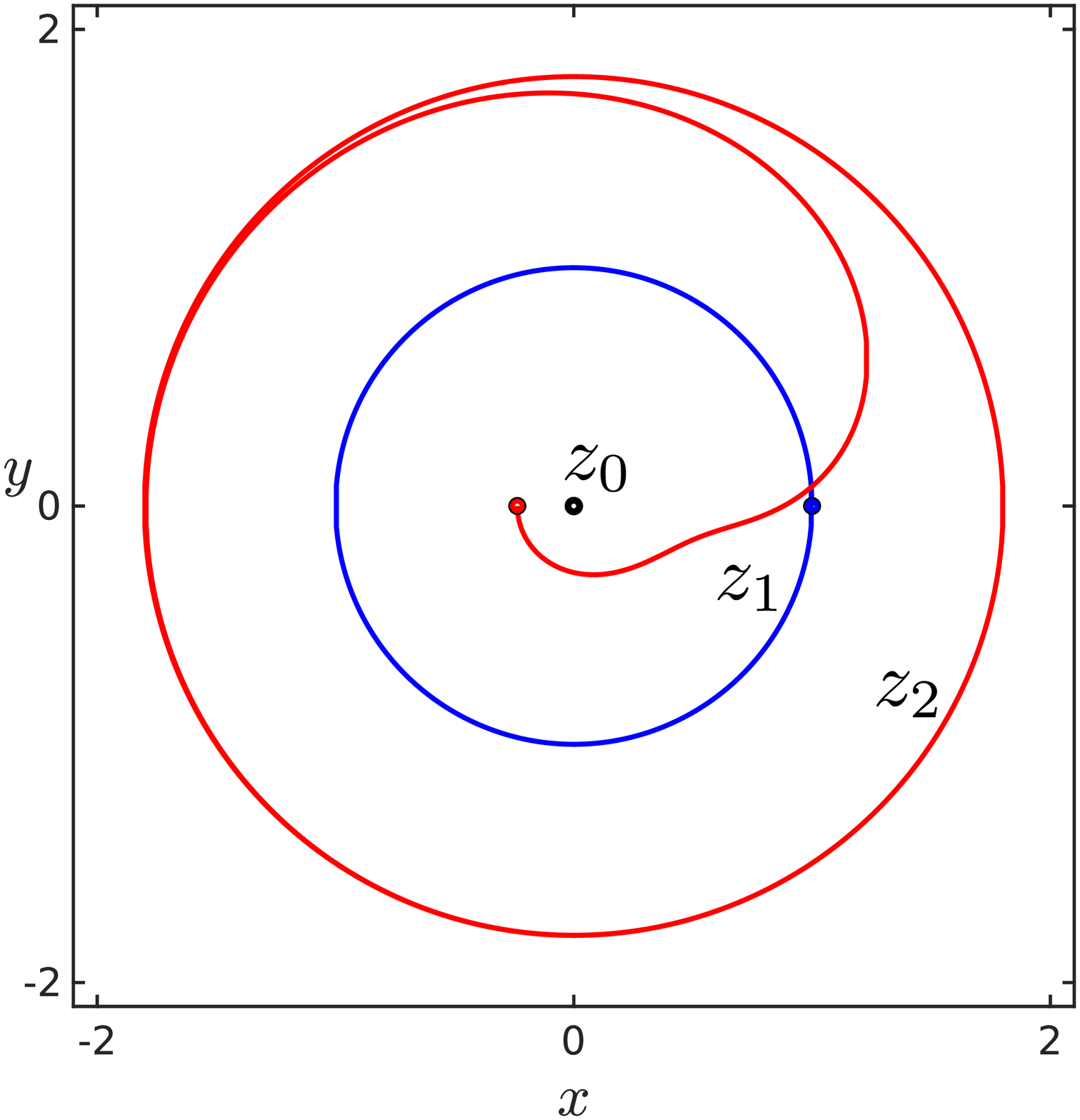}}
\label{figure_aperiodic_xy}
\caption{\label{figure_aperiodic}\small{ An example of an aperiodic vortex motion which is not a fixed configuration ($\Gamma_0=\Gamma_1=10,\Gamma_2=0,z_1|_{t=0}=1,z_2|_{t=0}=\eta_2|_{t=0}\approx-0.236946 ,z_0=0$). The positions of the vortices are shown in the (a) rotating and (b) stationary frame of references. The vortex $\mathcal{V}_2$ (red trajectory) asymptotically approaches the fixed configuration trajectory described in figure~\ref{figure_fixed_conf}.} }
\end{figure}
Any initial condition that corresponds to a point on a homoclinic orbit in the $(u,v)$ phase plane yields this type of vortex motion.  Since a trajectory that starts from  any such point, by definition,  asymptotically tends to a  saddle equilibrium point (which corresponds to a collinear fixed configuration in the physical plane), we expect the  vortex $\mathcal{V}_2$ to move  in a nearly circular path around $\mathcal{V}_0$ after a sufficiently long time. We have illustrated this in figure~\ref{figure_aperiodic}, by considering $\Gamma_0=10,\Gamma_1=10$ (i.e., $\alpha=1$), $z_1|_{t=0}=1$ and $\eta_2|_{t=0}\approx -0.236946$. The initial condition corresponds to the unique $u$-axis intersection point of the homoclinic orbit  that tends to $u_3$. The initial condition for $z_2$ was  calculated  by numerically solving $H(u,0)=H(u_3,0)$ to a very high working precision. As one can see in figure~\ref{figure_aperiodic}(a), the  $(u,v)$ phase-plane trajectory of $\eta_2$ tends to the equilibrium point $u_3\approx 1.80194$, whereas in the stationary frame of reference, see figure~\ref{figure_aperiodic}(b), the vortex $\mathcal{V}_2$ approaches the circular path, which was described earlier in figure~\ref{figure_fixed_conf}(b).

\clearpage
%-----------------------------------------
\subsection{Periodic inter-vortex distance case}
%-----------------------------------------

\begin{figure}[htbp]
\centering
\subfigure[]{
\includegraphics[height=2.2in]{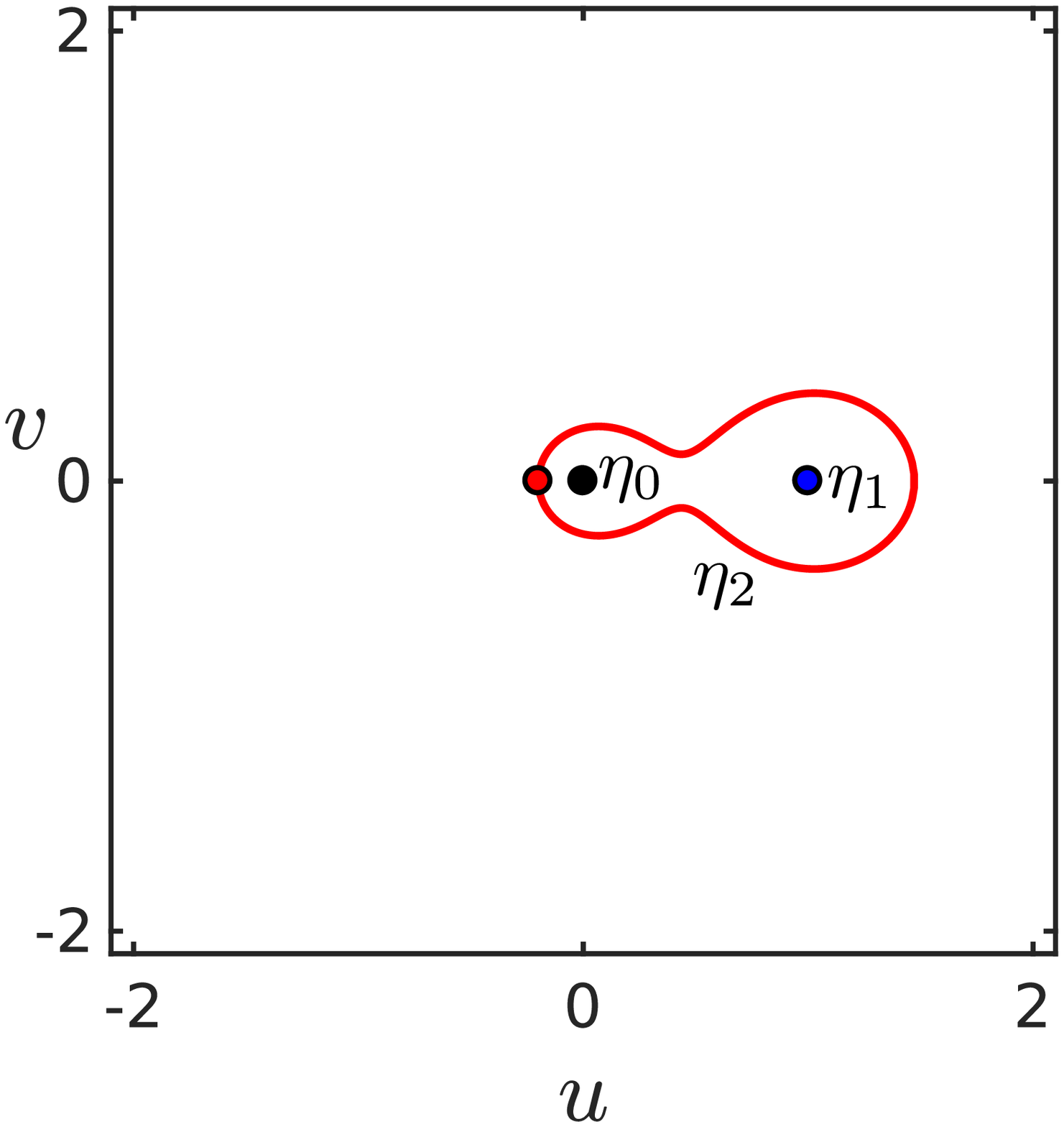}}\,
\label{figure_periodic_uv}
\subfigure[]{\includegraphics[height=2.2 in]{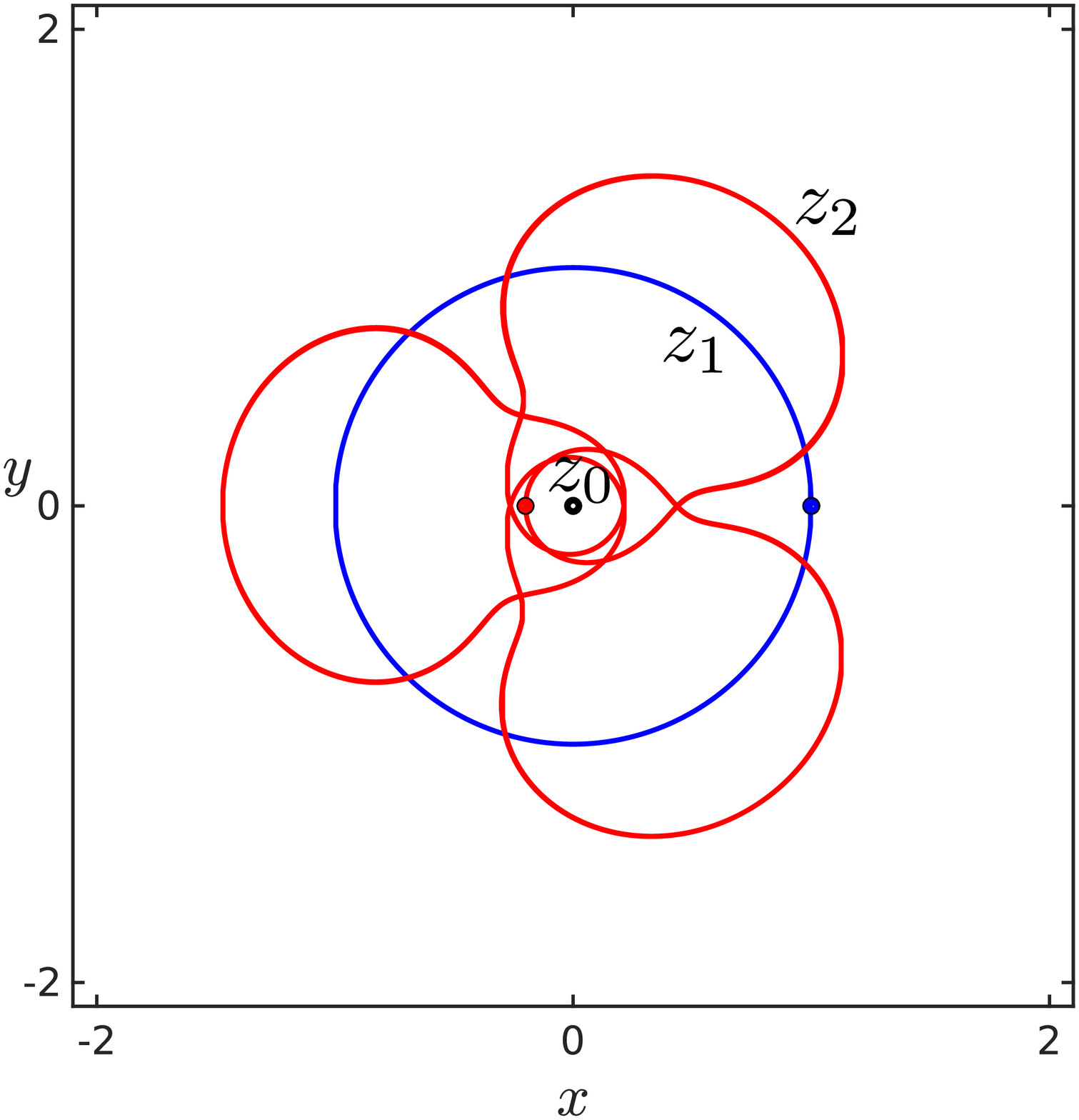}}\,
\label{figure_periodic_xy}
\subfigure[]{\includegraphics[height=2.2 in]{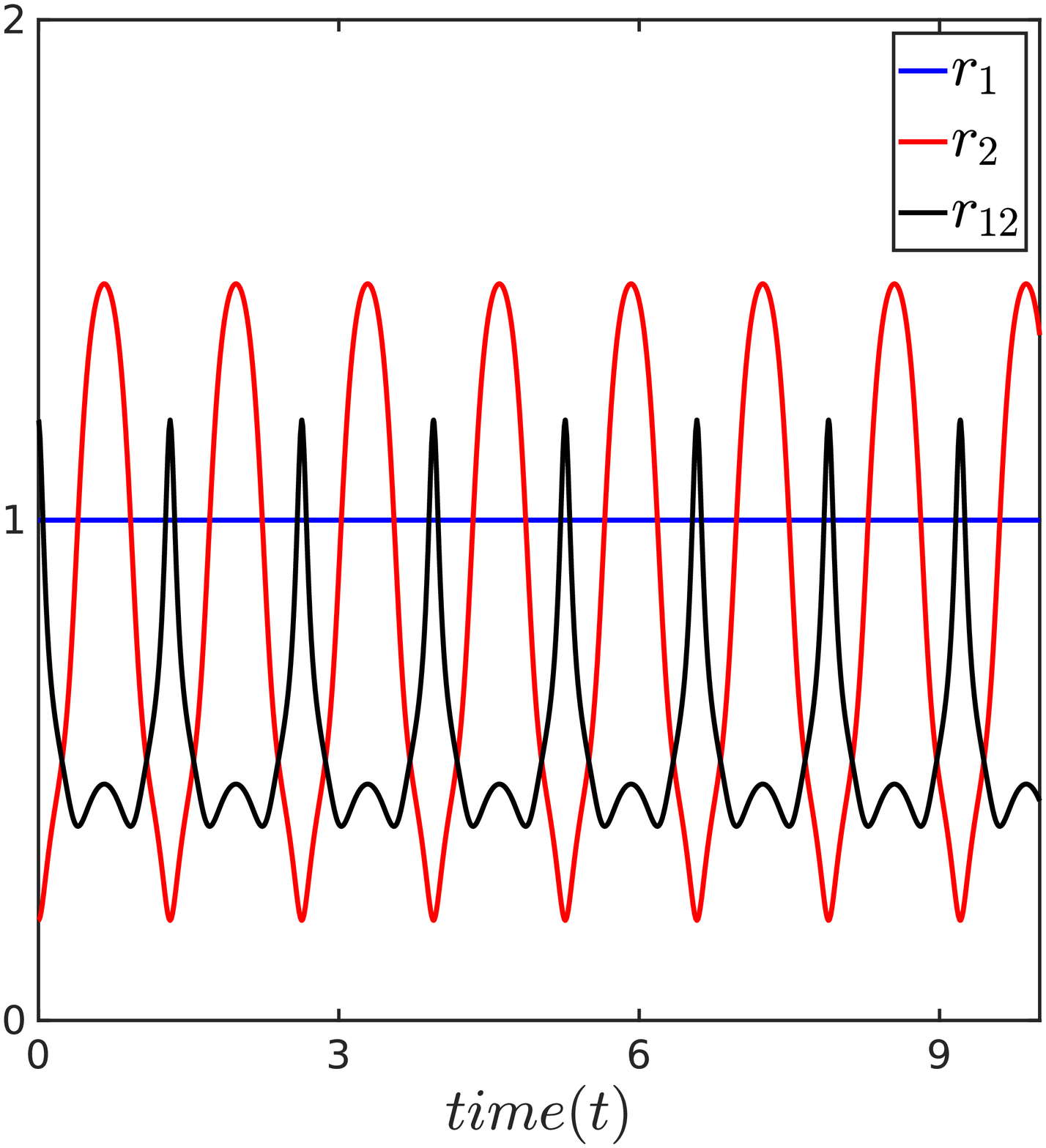}}\,
\label{figure_periodic_distances}
%\subfigure[]{\includegraphics[height=2.5in]{figures/periodic_distances.eps}}
%\label{figure_restricted_opposite}
\caption{\label{figure_periodic}\small{An example of a vortex motion in which the inter-vortex distances $r_2$ and $r_{12}$ are periodic in time ($\Gamma_0=\Gamma_1=10,\Gamma_2=0,z_1|_{t=0}=1,z_2|_{t=0}=-0.2,z_0=0$). The positions of the vortices are shown in the (a) rotating frame of reference (b) stationary frame of reference. (c) The inter vortex distance functions are plotted against time. $r_1$ is the constant unity function, whereas $r_2$ and $r_{12}$ are periodic.}  }
\end{figure}
All initial conditions that do not belong in the above two categories must lead to a vortex motion in which inter-vortex distances $r_2$ and $r_{12}$ are periodic. One such situation is depicted in figure~\ref{figure_periodic}  by considering $\Gamma_0=10,\Gamma_1=10, z_1|_{t=0}=1$ and $\eta_2|_{t=0}=-0.2$. As one would expect, we have a closed trajectory in the $(u,v)$ phase plane, as shown in figure~\ref{figure_periodic}(a), whereas the actual vortex trajectories look like as in figure~\ref{figure_periodic}(b). The periodicity in the variables $r_2$ and $r_{12}$ is evident from figure~\ref{figure_periodic}(c).

%-----------------------------------------
\section{Comparison with the Classical problem}
%-----------------------------------------

By the ``classical restricted three-vortex" problem---here onwards ``classical" problem---we mean our model without the added assumption that the vortex $\mathcal{V}_0$ is fixed in the plane.
In this section, we shall compare our results with those of the classical problem~\cite{AB2016}. 

As the vortex $\mathcal{V}_2$ has zero circulation, it no way affects the motion of the two-vortex system constituted by the vortices $\mathcal{V}_0$ and $\mathcal{V}_1$. Hence in the classical problem, depending on the value of the sum $\Gamma_0+\Gamma_1$, there are two possible scenarios: (i) if $\Gamma_0+\Gamma_1=0$, then the two vortices translate uniformly to infinity in the direction perpendicular to the line joining them, and (ii) if $\Gamma_0+\Gamma_1\neq0$, they exhibit a circular motion around the center of vorticity $(\Gamma_0z_0+\Gamma_1z_1)/(\Gamma_0+\Gamma_1)$ with a constant angular velocity~\cite{NP2013}. In both cases, the vortices $\mathcal{V}_0$ and $\mathcal{V}_1$ maintain a constant distance between them.

In the first case, i.e., $\Gamma_0+\Gamma_1=0$, it can be shown that, if the distance between the vortex $\mathcal{V}_2$ and the other two vortices are small enough initially, it is for all time~\cite{KC1989}. Consequently, the third vortex $\mathcal{V}_2$ can also exhibit an unbounded motion in the classical problem. On the contrary, the vortex pair $\mathcal{V}_0$ and $\mathcal{V}_1$ can no longer escape to infinity once vortex $\mathcal{V}_0$ is fixed as in the present problem, resulting in vortices being  entrapped in a neighbourhood of the origin irrespective of the vortex circulations (see, lemma~\ref{lemma_restricted_bound}).

Let us now consider the case $\Gamma_0+\Gamma_1\neq 0$. For any such pair of circulations $(\Gamma_0,\Gamma_1)$, there are two relative equilibria which are in equilateral-triangle configurations. Therefore, unlike the present problem (i.e.,~$\mathcal{V}_0$ is fixed), the classical problem (i.e.,~$\mathcal{V}_0$ is free) permits fixed configurations which are not collinear. Although the number of the collinear fixed configurations remains the same (three for $\alpha=\Gamma_1/\Gamma_0>0$ and one for $\alpha=\Gamma_1/\Gamma_0<0$), there is a change in the stability of such configurations. In the classical problem, all three of them are unstable for $\alpha>0$, whereas the fixed vortex variant allows a neutrally stable one. For $\alpha<0$, it is neutrally stable in the classical problem and unstable in the fixed vortex model. 
%
%These observations indicate that introducing constraints such as fixing vortex at a suitable location in the fluid system can potentially change the boundedness of vortices, reduce the number of fixed configurations of vortices and their stability, thereby enhancing properties such as  mixing.
\section{Conclusions}
The different types of motion exhibited by a pair of point vortices (vortex $\mathcal{V}_1$ with a non-zero circulation $\Gamma_1$, and vortex $\mathcal{V}_2$ with zero circulation) in the presence of the velocity field induced by a fixed point vortex (vortex $\mathcal{V}_0$ with a non-zero circulation $\Gamma_0$), in an ideal two-dimensional incompressible flow has been analysed in detail. Although vortex $\mathcal{V}_1$ executes a simple circular motion around the fixed vortex $\mathcal{V}_0$ with constant angular velocity $\omega=\Gamma_0/2\pi$, the vortex $\mathcal{V}_2$ can have a wide range of complex trajectories that deviate significantly from a closed circular path. Nevertheless, if one looks at the motion of vortex $\mathcal{V}_2$ from a rotating frame of reference within which the vortex $\mathcal{V}_1$ is stationary, it turns out that the underlying dynamical equations have a Hamiltonian structure and thus the trajectories in this new coordinate system are simply the level curves of the  Hamiltonian~\eqref{Hamiltonian_restricted}.

There are several advantages of studying the vortex motion in the above mentioned rotating frame of reference. For example, one could use the Hamiltonian structure of system~\eqref{udotvdot_restricted} along with the fact that the inter-vortex distances remain invariant under the coordinate transformation~\eqref{eq_coordinate_transform} to quickly establish that the vortex motion is  always bounded [lemma~\ref{lemma_restricted_bound}].  Furthermore, this formulation enables us to  classify all possible vortex motion into three categories since a trajectory in the $(u,v)$ phase plane is either (i) an equilibrium solution, (ii) a closed trajectory, or (iii) a homoclinic orbit which physically  correspond to  (i) a fixed configuration of vortices, (ii) a vortex motion in which the inter-vortex distances are periodic functions of time with the vortices oscillating  between two distinct collinear configurations, or (iii) a vortex motion in which the three vortex system asymptotically tends to a collinear fixed configuration, respectively [theorem~\ref{therorem_trajectories_uv}].

The idea of using a rotating reference frame is naturally suited for studying the fixed configurations of vortices, as it simplifies the process from looking for special solutions of the complex dynamical system~\eqref{z0_eqn}--\eqref{z2_eqn} to finding the equilibrium solutions of a relatively easier Hamiltonian system~\eqref{udotvdot_restricted}. Adopting the above approach, it has been shown that the restricted three-vortex system~\eqref{z0_eqn}--\eqref{z2_eqn} is in a fixed configuration if and only if the vortex $\mathcal{V}_2$ initially lies along the line joining vortices $\mathcal{V}_0$ and $\mathcal{V}_1$ at any one of the real roots of the cubic polynomial \eqref{u_fixedpoint_cubic_polynomial}. Hence, in all fixed configurations, vortex $\mathcal{V}_2$  moves in a circular path around the fixed vortex, maintaining collinearity with vortices $\mathcal{V}_0$ and $\mathcal{V}_1$. 
%Rather than  looking for  special solutions of the complex system~\eqref{z0_eqn}--\eqref{z2_eqn}, it simplifies the process to finding  and analysing the equilibrium solutions of a relatively easier Hamiltonian system~\eqref{udotvdot_restricted}.
% Adopting the above approach, it has been shown that in all fixed configurations, vortex $\mathcal{V}_2$  moves in a circular path around the fixed vortex, maintaining collinearity with vortices $\mathcal{V}_0$ and $\mathcal{V}_1$. Moreover, the necessary and sufficient condition for a fixed configuration has been found to be 
% The necessary and sufficient condition for any such configuration is found to be $\mathcal{V}_2$ being positioned along the line joining $\mathcal{V}_0$ and $\mathcal{V}_1$ at any one of the real roots of the cubic polynomial \eqref{u_fixedpoint_cubic_polynomial}. 
If $\Gamma_0\Gamma_1>0$, there are three distinct fixed collinear configurations, with two being unstable and one neutrally stable, whereas if $\Gamma_0\Gamma_1<0$, there is only one unstable fixed collinear configuration [lemma~\ref{lemma_restricted_equilibrium_points_position} and \ref{lemma_restricted_saddles_centers_numbers}, and proposition~\ref{prop_fixed_config}].

The presence of a fixed vortex in a vortex system appears to change its dynamics significantly. Even a seemingly simple vortex system such as the restricted three-vortex system \eqref{z0_eqn}--\eqref{z2_eqn} indicates that introducing a fixed vortex at a suitable location in the fluid can potentially change the boundedness of vortices, reduce the number of fixed configurations of vortices and its stability, thereby enhancing physical properties such as  heat and mass transport, mixing etc. Hence from a  theoretical and application viewpoints, the effects of fixed vortices in a general $N$-vortex system is worth exploring further. The authors also believe that the dynamical system theory approach presented in this paper will be useful for studying and classifying different special vortex motions like self-similar evolutions and relative equilibria which are of physical importance.

%The presence of a fixed vortex in the system appears to change its dynamics significantly. While the classical restricted three-vortex problem permits unbounded vortex motion for a counter-rotating pair case (i.e., $\Gamma_0+\Gamma_1=0$), the same with a fixed vortex does not allow any. Apart from the collinear fixed configurations, the classical problem has two equilateral triangle fixed configurations for any pair of circulations $(\Gamma_0,\Gamma_1)$, which are absent once vortex $\mathcal{V}_0$ is fixed. Interestingly, there is also changes in the stability properties of the fixed collinear configurations. The unique collinear fixed configuration in the $\Gamma_0\Gamma_1<0$ case is stable in the classical problem, whereas unstable in the fixed vortex variant. Moreover, for $\Gamma_0\Gamma_1>0$ all three collinear fixed configurations are unstable in the classical problem, but the fixed vortex variant allows one neutrally stable configuration. 

%
%By selectively placing a finite number of fixed vortices on the plane In situations, where one would like to tweak the trajectories of fluid particles    

\section*{Acknowledgment}
P.S. acknowledges financial support from IIT Madras
through the Grant No. MAT/16-17/671/NFSC/PRIY. We thank Prof.~Shaligram Tiwari, IIT Madras, for fruitful discussions.

\appendix
\section{Index Theory}
\label{appendix:index}
Here, we will slightly modify the definitions of the index and state the resulting properties, accommodating the singularity points $\eta_0$ and $\eta_1$ of system~\eqref{udotvdot_restricted}. We have omitted the proofs since it follows the same lines as in~\citet{PL2013}. Let us begin by introducing the following  new terminology.
% Let us begin by defining the index of a Jordan curve (a piecewise smooth, simple closed curve) contained in the phase plane.
%
\begin{definition}{(Non-regular point)}
A point in the phase plane is said to be non-regular, if it is either an equilibrium or a singularity point of system~\eqref{udotvdot_restricted}. 
\end{definition}
We shall now define the index of a Jordan curve (a piecewise smooth, simple closed curve) contained in the $(u,v)$ phase plane.
\begin{definition}{(Index of a Jordan curve)} Let $\gamma:[0,1]\to \mathbb{R}^2$ be a Jordan curve parametrized in the anti-clockwise direction and suppose that $\gamma([0,1])$ does not contain any of the non-regular points of system $\eqref{udotvdot_restricted}$. Note that the angle $\phi$, the field $\mathbf{F}=(\dot{u},\dot{v})$ makes with the $u$-axis is $\tan^{-1}(\dot{v}/\dot{u})$, and it is well defined and piecewise smooth when restricted to $\gamma([0,1])$. The \emph{index $I_F(\gamma)$} of $\gamma$ is then defined as 
\begin{equation}
I_F(\gamma)=\frac{1}{2\pi}\oint_\gamma d\phi,
\end{equation}
the total number of anti-clockwise revolutions made by the field $F$ along $\gamma$ in one circuit.
\end{definition}
Let us now look at some of the key properties of index.
\begin{lemma}
\begin{figure}[!htbp]
\centering
\includegraphics[height=1in]{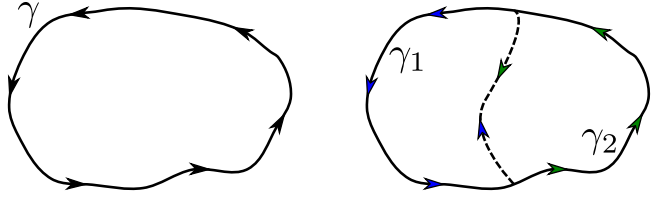}
\caption{\label{figure_jordan_curve} The Jordan curve  $\gamma$ is decomposed into two other Jordan curves $\gamma_1$ and $\gamma_2$.}
\end{figure}
\label{lemma_index_sum_curves}
 If a Jordan curve $\gamma$ can be broken down into two Jordan curves $\gamma_1$ and $\gamma_2$ (see figure~\ref{figure_jordan_curve}) then 
\begin{equation}
I_F(\gamma)=I_F(\gamma_1)+I_F(\gamma_2)
\end{equation}
\end{lemma}
\begin{theorem} 
\label{theorem_index_zero}
If a Jordan curve $\gamma$  does not contain any non-regular points of system~\eqref{udotvdot_restricted} on it or its interior, then $I_F(\gamma)=0$.
\end{theorem}
\begin{definition}{(Index of a non-regular point)}
 Let $\gamma$ be a Jordan curve that contains exactly one non-regular point $\mathbf{x^*}$ of system~\eqref{udotvdot_restricted} in  its interior. The index of $\mathbf{x^*}$, denoted by $I_F(\mathbf{x^*})$ is then defined as 
 \begin{equation}
 I_F(\mathbf{x^*})=I_F(\gamma)
 \end{equation}
\end{definition}
The above definition is well defined. Suppose  $\gamma_1$ and $\gamma_2$ are  two different Jordan curves, both enclosing and only enclosing a non-regular point $\mathbf{x^*}$ in their interior. We need to show that $I_F(\gamma_1)=I_F(\gamma_2)$. Let us first consider the case when one of the curve is contained in the interior of the other. WLOG let us assume $\gamma_1$ is contained in the interior of $\gamma_2$. Consequently, there are no non-regular points in the intersection of interior of $\gamma_2$ and exterior of $\gamma_1$. Therefore, one could continuously deform $\gamma_2$ into $\gamma_1$ without encountering any non-regular points. The index must also change continuously during the process. Since the index is integer valued, continuity would essentially mean that it is a constant, i.e.,  $I_F(\gamma_1)=I_F(\gamma_2)$. In a general situation, one could consider a circle $\gamma$ centered at $\mathbf{x^*}$ with sufficiently small radius and contained in the intersection of interior of $\gamma_1$ and $\gamma_2$. The above arguments repeated for $\gamma_1$ and $\gamma$,  $\gamma_2$ and $\gamma$ would yield us $I_F(\gamma_1)=I_F(\gamma)$ and  $I_F(\gamma_2)=I_F(\gamma)$, respectively. Therefore, $I_F(\gamma_1)=I_F(\gamma_2)$ for any two Jordan curves that contain only one non-regular point $x^*$  in their interior.       
\begin{theorem}
\label{theorem_app:sum_index}
Let $\gamma$ be a Jordan curve in the phase plane and suppose there are a total of $n$ non-regular points, $\mathbf{x_i^*}$, $i=1,2,..,n$ in the interior of $\gamma$, then  
\begin{equation}
I_F(\gamma)=\sum_{i=1}^nI_F(\mathbf{x^*_i})
\end{equation}
\end{theorem}
\begin{theorem}
\label{theorem_app:closed_trajectory}
If $\gamma$ is a closed trajectory of system~\eqref{udotvdot_restricted}, then $I_F(\gamma)=1$.
\end{theorem}

\clearpage
\bibliography{references.bib}% Produces the bibliography via BibTeX.

\end{document}